\documentclass[pra,aps,twocolumn,groupedaddress,nofootinbib]{revtex4-2} %superscriptaddress
\usepackage{graphicx}
\usepackage[nointegrals]{wasysym} %
\usepackage[export]{adjustbox}
\usepackage{amsmath,amsfonts,amssymb,latexsym}
\usepackage{hhline}
\usepackage{bm}
\usepackage{verbatim}
\usepackage{enumitem}
\usepackage{mathrsfs}
\usepackage{slashed}
\usepackage{empheq}
\usepackage{physics}

\usepackage{graphicx}
\usepackage{dcolumn}
\usepackage{bm}
\usepackage{multirow}

\usepackage[normalem]{ulem}

\usepackage{amsmath}
\usepackage{amsthm}
\usepackage{amstext}
\usepackage{amssymb}
\usepackage{mathrsfs}
\usepackage{amsfonts}
\usepackage{amsbsy} 

\usepackage{braket}

\usepackage[all,matrix,cmtip]{xy}
\usepackage{mathtools}

\usepackage{xcolor}
\definecolor{red}{rgb}{1,0,0}
\definecolor{blue}{rgb}{0,0,1}
\definecolor{dblue}{rgb}{0,0,0.4}
\definecolor{green}{rgb}{0,1,0}
\definecolor{black}{rgb}{0,0,0}
\definecolor{white}{rgb}{1,1,1}
\definecolor{pastelblue}{RGB}{20,93,160}

\definecolor{brn}{rgb}{.8,.4,.0}
\definecolor{redo}{rgb}{1,.5,.0}
\definecolor{ddgrn}{rgb}{0,0.4,0}
\definecolor{dgrn}{rgb}{0,0.55,0}
\definecolor{dbl}{rgb}{0,0,0.5}

\usepackage[colorlinks,citecolor=pastelblue,linkcolor=pastelblue,urlcolor=pastelblue]{hyperref}

\newcommand{\sgn}{{\rm sgn}}

\newcommand{\bpm}{\begin{pmatrix}}
	\newcommand{\epm}{\end{pmatrix}}
\newcommand{\bmm}{\begin{matrix}}
	\newcommand{\emm}{\end{matrix}}
\newcommand{\bvm}{\begin{vmatrix}}
	\newcommand{\evm}{\end{vmatrix}}

\usepackage{euscript}

\makeatletter
\newsavebox{\@brx}
\newcommand{\llangle}[1][]{\savebox{\@brx}{\(\m@th{#1\langle}\)}%
	\mathopen{\copy\@brx\kern-0.5\wd\@brx\usebox{\@brx}}}
\newcommand{\rrangle}[1][]{\savebox{\@brx}{\(\m@th{#1\rangle}\)}%
	\mathclose{\copy\@brx\kern-0.5\wd\@brx\usebox{\@brx}}}
\makeatother

\allowdisplaybreaks

\newcommand{\bs}{\boldsymbol}

\begin{document}

%\preprint{APS/123-QED}

\title{Non-Abelian topological superconductivity \\ from melting Abelian fractional Chern insulators}% Force line breaks with \\

\author{Zhengyan Darius Shi}%
\email{zhengyanshi@stanford.edu}
\affiliation{
Department of Physics, Stanford University, Stanford, California 94305, USA}

\author{T. Senthil}
\email{senthil@mit.edu}
 %\homepage{}
\affiliation{
Department of Physics, Massachusetts Institute of Technology,
Cambridge, Massachusetts 02139, USA
}%

\date{\today}

\begin{abstract}
    Fractional Chern insulators (FCI) are exotic phases of matter realized at partial filling of a Chern band that host fractionally charged anyon excitations. Recent numerical studies in several microscopic models reveal that increasing the bandwidth in an FCI can drive a direct transition into a charge-2e superconductor rather than a conventional Fermi liquid. Motivated by this surprising observation, we propose a theoretical framework that captures the intertwinement between superconductivity and fractionalization in a lattice setting. Leveraging the duality between three field-theoretic descriptions of the Jain topological order, we find that bandwidth tuning can drive a single parent FCI at $\nu = 2/3$ into five different superconductors, some of which are intrinsically non-Abelian and support Majorana zero modes. %The quantum critical points associated with these transitions provide condensed matter realizations of gapless non-Abelian Chern-Simons matter theories. 
    Our results reveal a rich landscape of exotic superconductors with no normal state Fermi surface and predict novel higher-charge superconductors coexisting with neutral non-Abelian topological order at more general filling fractions $\nu = p/(2p+1)$. 
\end{abstract}

\maketitle

\section{Introduction}

Superconductivity (SC) and the fractional quantum Hall (FQH) effect are two of the most striking emergent phenomena in condensed matter physics. Historically, research in these two directions developed largely in parallel due to the apparent incompatibility between BCS pairing and the strong magnetic fields that stabilize FQH states. However, the discovery of fractional quantum anomalous Hall phases in two-dimensional Van der Waals materials~\cite{Cai2023_FQAHTMD,Park2023_FQAH_TMD,Xu2023_FQAHTMD,Zeng2023_FQAHTMD,Lu2023_FQAHPenta} at zero external magnetic field challenges this conventional wisdom. Moreover, recent transport measurements show signatures of both SC and FQH phases in neighboring regions of the phase diagram of twisted MoTe$_2$~\cite{Xu2025_SCdopeTMD} as well as rhombohedral graphene~\cite{Han2024_chiralSC_penta}. These experimental advances raise an interesting theoretical question: what is the relationship between SC and FQH and how do we describe their intertwinement in a unified theoretical framework? 

During the past year, a growing body of work has explored the interplay between SC and FQH along the doping axis~\cite{Shi2024_doping,Kim2024_anyonSC,Shi2025_dopeMR,Shi2025_anyon_delocalization,Nosov2025_plateau,Pichler2025_anyonSC,Zhang2025_SU(3)1_dope,Nakajima2025_thermo_anyon}, building on the idea of ``anyon superconductivity" proposed by Laughlin in the context of high-$T_c$ superconductors~\cite{Laughlin1988_anyonSC,Fetter1989_anyonSC_RPA, Lee1989_anyonSC,Chen1989_anyonSC}.\footnote{See Refs.~\cite{Divic2024_HofHubb,Kuhlenkamp2025_HofHubb,Han2025_anyonexciton} for related works in different settings.} Starting with a fractional Chern insulator (FCI) on the lattice (at zero or nonzero external field)~\cite{Sun2011_FCI,Sheng2011_FCI,Regnault2011_FCI,Tang2011_FQAH,Neupert2010_FQAH}, doping introduces a finite density of anyons with fractional charge and fractional statistics. As emphasized in Ref.~\cite{Shi2024_doping}, exact lattice translation symmetry endows these anyons with nontrivial band dispersion (see microscopic calculations in Refs.~\cite{Goncalves2025_anyondisp,Schleith2025_anyondisp,Yan2025_anyondisp}). Under suitable energetic assumptions, the interactions between anyons induce superconductivity despite strong time-reversal symmetry breaking and the absence of a Fermi surface in the high temperature normal state.

In this letter, we focus on an axis in the phase diagram orthogonal to the doping axis. Specifically, starting with an FCI realized at partial filling of some topological band, we ask whether tuning the bandwidth can drive a direct transition into a superconductor. In bosonic FCIs, the additional kinetic energy from the increased bandwidth naturally favors a superfluid phase. Indeed, at lattice filling $\nu = 1/2$, a direct continuous transition between the simplest bosonic Laughlin state $U(1)_2$ and a topologically trivial superfluid was proposed theoretically~\cite{Barkeshli2012_LaughlinSF,Song2023_QPT_FQAH} and later confirmed via density-matrix renormalization group (DMRG) simulations~\cite{Wang2025_LaughlinSF}. %This transition provides a condensed matter realization of a strongly coupled 2+1 dimensional conformal field theory known as $N_f = 2$ QED-CS, in which two flavors of emergent Dirac fermions interact with a gapless $U(1)$ gauge field.

\begin{figure}
    \centering
    \includegraphics[width=0.9\linewidth]{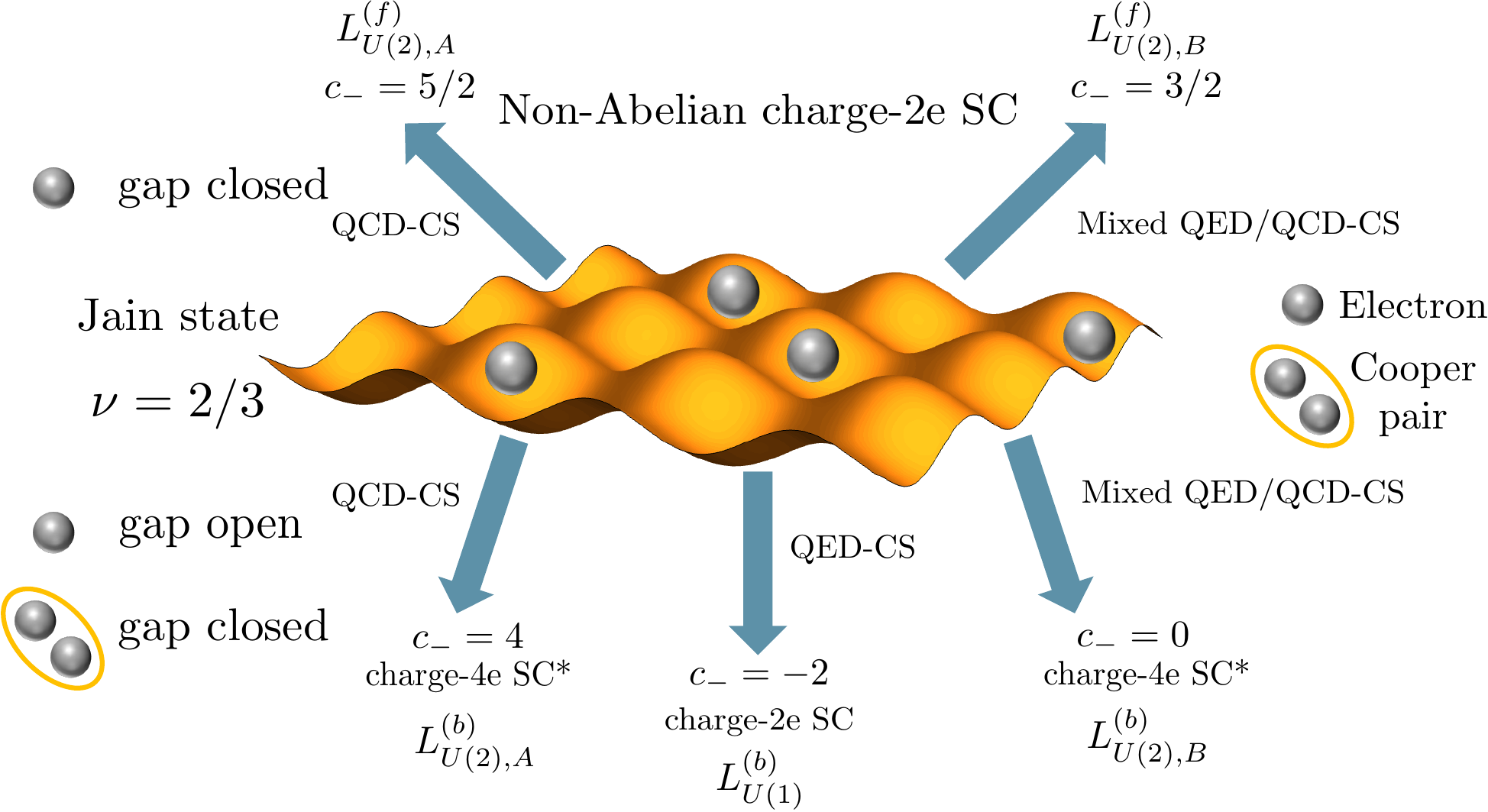}
    \caption{Five distinct superconductors that can be accessed from the Jain state at $\nu = 2/3$ through a direct bandwidth-tuned quantum phase transition. Note that closing/not closing the electron gap leads to superconductors with half-integer/integer chiral central charge $c_-$. }
    \label{fig:Jain23_SC}
\end{figure}

The generalization from bosonic to fermionic FCIs presents new challenges. With broken time reversal symmetry, fermions with a large bandwidth prefer to form an ordinary Landau Fermi liquid with no weak-coupling BCS instability. Therefore, it is natural to expect that increasing the bandwidth drives a direct transition from a fermionic FCI to a Fermi liquid metal. Surprisingly, a recent DMRG study of electrons at $\nu = 2/3$ finds a two-stage transition in which a chiral superconductor emerges at intermediate bandwidth, in between the FCI and the Fermi liquid~\cite{Wang2025_SCmeltFCI}.\footnote{Chiral topological SC neighboring an FCI in an ideal Chern band has also been found in Refs.~\cite{Guerci2025_FCISC,Guerci2026_TSC_vorlat} through exact diagonalization.} Moreover, since the order parameter has odd angular momentum, the SC likely has odd chiral central charge and hosts non-Abelian Majorana zero modes in its vortex cores~\cite{Read1999_pair, Ivanov2001_nonabelian}. 

\begin{figure}
    \centering
    \includegraphics[width=0.9\linewidth]{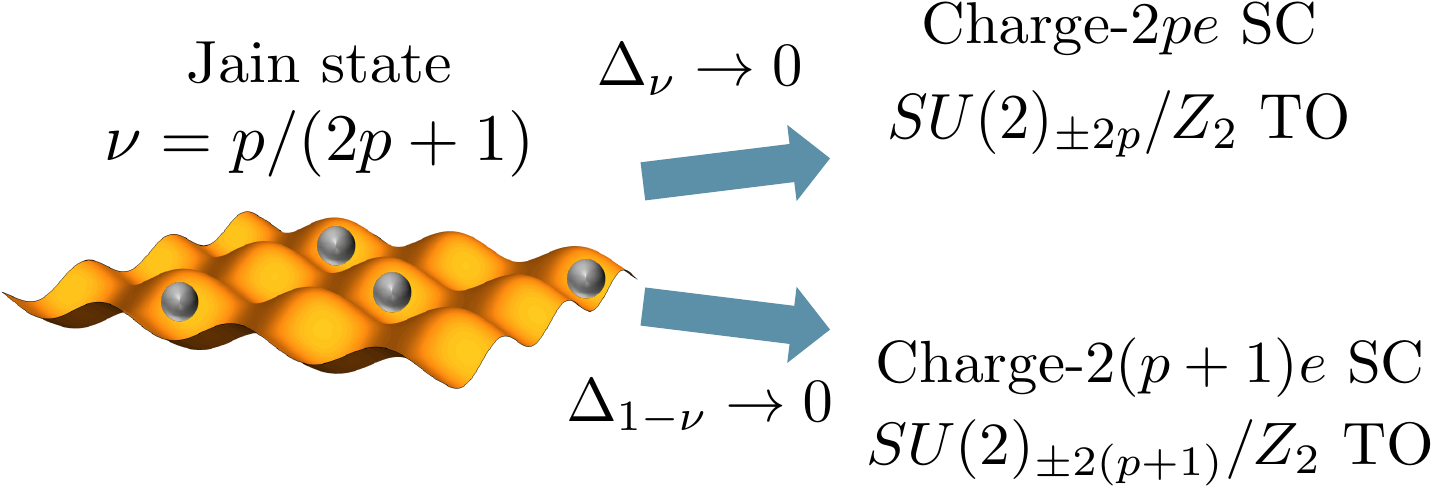}
    \caption{For $p \neq 1, -2$, we find four higher-charge superconductors accessible from the Jain FCI at $\nu = p/(2p+1)$ through a bandwidth-tuned phase transition. $\Delta_q$ is the energy gap of the anyon with fractional charge $q$ and the two families of transitions involve gap closing of $\Delta_{\nu}$ and $\Delta_{1-\nu}$ respectively.}
    \label{fig:arbitraryJain_SC}
\end{figure}

Motivated by this intriguing numerical observation, we focus on lattice filling $\nu = 2/3$ and develop a theory of bandwidth-tuned transitions from the Jain FCI to five superconductors with distinct topological properties, as summarized in Fig.~\ref{fig:Jain23_SC}. In three of these transitions, the electron gap remains open and the resulting phase is either a charge-2e SC with $c_- = -2$ or a charge-4e SC* with $c_- = 4$/$c_- = 0$ and a coexistent Abelian topological order. In the other two transitions, the electron gap closes and the resulting phase is a charge-2e topological SC with $c_- = 3/2$ or $c_- = 5/2$. The technical tools we develop for $\nu = 2/3$ generalize to arbitrary Jain states at filling fractions $\nu = p/(2p+1)$, whose neighboring superconductors generically carry a non-Abelian topological order. The full landscape of Jain-SC transitions is shown in Fig.~\ref{fig:arbitraryJain_SC}. 

%\ZS{Added paragraph here about relationship to doping.}
Following the general framework of Ref.~\cite{Shi2024_doping}, we can show that all of the new SCs in Fig.~\ref{fig:Jain23_SC} and Fig.~\ref{fig:arbitraryJain_SC} can also be accessed by doping the corresponding Jain FCI. These doping-induced SCs vastly generalize the examples constructed in Ref.~\cite{Shi2024_doping,Shi2025_dopeMR}.

\begin{comment}
Outline
\begin{enumerate}
    \item Begin with broad motivation: relationship between SC and FQH, two of the most remarkable emergent phenomena in condensed matter physics. Explain why they are conventionally understood to be mutually exclusive. But discovery of FQAH invites us to revisit the story. 
    \item Simplest example: bosonic Laughlin-SF transition driven by bandwidth tuning. Discuss early theoretical work by Barkeshli McGreevy and more recent numerics. 
    \item From boson to fermion: while SF is a natural conducting state of bosons favored by bandwidth, fermions with large bandwidth like to form Fermi surfaces and superconductivity only comes as a secondary instability. Mention Taige's numerics as a surprise that calls for understanding. 
    \item Explain strategy: (1) focus on the Jain state at $\nu = 2/3$ and explain there are two paths towards a SC (need a figure). One path involves closing the boson gap without the fermion gap. All the action is happening in the bosonic sector and transition theory quite similar to the bosonic Laughlin-SF transition. The second path involves closing the fermion gap. This is more surprising because fermions need to avoid forming a FL. Nevertheless, we find a transition described in terms of a more exotic QCD-CS theory with non-Abelian gauge fields. Resulting SC is naturally topological, with vortices trapping Majorana zero modes. 
    \item Explain generalization to all the Jain states (another figure). Perhaps comment on agreement/disagreement with numerics? A bit delicate...
\end{enumerate}
\end{comment}

\section{Three descriptions of the Jain FCI at \texorpdfstring{$\nu = 2/3$}{}}

The key observation underlying the multiple transition pathways in Fig.~\ref{fig:Jain23_SC} is that the Jain FCI at $\nu = 2/3$ (henceforth referred to as $\mathrm{Jain}_{2/3}$) admits three distinct Lagrangian descriptions, which we now derive.

Consider electrons at lattice filling $\nu = 2/3$ with $2\pi \mathbb{Z}$ magnetic flux per unit cell. At each lattice site $\bs{r}$, we decompose the electron operator $c(\bs{r})$ as a product of three partons $c(\bs{r}) = f_1(\bs{r}) f_2(\bs{r}) f_3(\bs{r})$ and assign electric charge $+1,+1, -1$ to $f_1, f_2, f_3$ so that $c$ carries the correct electric charge $+1$. %This decomposition introduces a local $SU(3)$ gauge redundancy under which 
%\begin{equation}
%    f_i \rightarrow \sum_{j=1}^3 U_{ij} f_j \,, \quad c \rightarrow c \, \det U = c \,, \quad U \in SU(3) \,. 
%\end{equation}
%A faithful representation of the original model can be obtained by coupling $f_i$ to a dynamical $SU(3)$ gauge field.
To construct the $\mathrm{Jain}_{2/3}$ state, we consider a mean-field ansatz in which each $f_i$ goes into a Chern insulator with Chern number $C_1 = C_2 = 1, C_3 = -2$~\cite{Jain1989_CFframework,Kol1993_Jainlattice,Sohal2017_fluxattach_FCI}. If the ansatz for $f_1, f_2$ are exactly identical, then the mean-field preserves a $U(2)$ gauge redundancy
\begin{equation}\label{eq:U(2)_redundancy}
    f_{i=1,2} \rightarrow \sum_{j=1}^2 \tilde U_{ij} f_j \,, \quad f_3 \rightarrow f_3 \det \tilde U^{-1} \,, \quad \tilde U \in U(2) \,. 
\end{equation}
The action of $\tilde U$ implies that $f_1, f_2$ transform in the $(1/2,1)$ representation of $U(2)$ while $f_3$ transforms in the $(0,-2)$ representation of $U(2)$.\footnote{Recall that $U(2) = SU(2) \times U(1)/\mathbb{Z}_2$. Irreducible representations of $U(2)$ are thus labeled by a pair $(j,n)$ where $j$ is the $SU(2)$ spin and $n$ is the $U(1)$ charge. The $\mathbb{Z}_2$ quotient gives a gluing condition $j + n/2 \in \mathbb{Z}$.} The low energy effective Lagrangian thus takes the general form
\begin{equation}\label{eq:U(2)general}
    L = L_{(1/2,1)}[f_i, \alpha + A I] + L_{(0,-2)}[f_3, - \Tr \alpha - A] \,,
\end{equation}
where $A$ is the background electromagnetic gauge field and $L_{(j,n)}$ describes a matter field in the $(j,n)$ representation minimally coupled to the emergent $U(2)$ gauge field $\alpha$. At energies below the gaps of $f_i$, we can integrate out the gapped partons and obtain the TQFT
\begin{equation}\label{eq:JainFCI_U(2)+}
    \begin{aligned}
    &L^{(+)}_{U(2),2/3} = \frac{1}{4\pi} \Tr \left[(\alpha + A I) d (\alpha + A I) + \frac{2}{3} \alpha^3\right] \\
    &\hspace{0.5cm} - \frac{2}{4\pi} (\Tr \alpha + A) d (\Tr \alpha + A) \\
    &= \frac{1}{4\pi} \Tr \left[\alpha d \alpha + \frac{2}{3} \alpha^3\right] - \frac{2}{4\pi} \Tr \alpha d \Tr \alpha - \frac{1}{2\pi} \Tr \alpha \, d A \,. 
    \end{aligned}
\end{equation}
By decomposing the $U(2)$ gauge field into $SU(2)$ and $U(1)$ components, we see that the above Lagrangian has a $SU(2)$ Chern-Simons term with level $= - 1$ and a $U(1)$ Chern-Simons term with level $= 6$. Following the standard convention, we refer to this theory as $U(2)_{-1,6}$. To confirm that this unfamiliar $U(2)$ gauge theory describes $\mathrm{Jain}_{2/3}$, we recall from Ref.~\cite{Cheng2025_orderingqh} that $\mathrm{Jain}_{2/3}$ is the unique fermionic topological order with Hall conductance $\sigma_{xy} = 2/3$, chiral central charge $c_- = 0$, and a three-fold torus ground state degeneracy. In SM Sec.~\ref{app:JainTQFT}, we check that all of these properties are satisfied by $U(2)_{-1,6}$, thereby confirming its equivalence to $\mathrm{Jain}_{2/3}$ (see Ref.~\cite{Ma2020_QCDtransition} for applications of the $U(2)$ gauge theory to other FCI phases and transitions). 

A distinct $U(2)$ gauge theory description of the same TQFT can be obtained by regarding $\mathrm{Jain}_{2/3}$ as an integer quantum Hall (IQH) state at $\nu = 1$ stacked with a hole Laughlin state at $\nu = -1/3$. The same decomposition $c = f_1 f_2 f_3$ applied to the $\nu = -1/3$ model gives
\begin{equation}
    L = L_{(1/2,1)}[f_i, \alpha+A] + L_{(0,-2)}[f_3, -\Tr \alpha - A] + \mathrm{CS}[A,g] \,, 
\end{equation}
where $\mathrm{CS}[A,g] = 1/(4\pi) A d A + 2 \mathrm{CS}_g$ is the response theory of the $\nu = 1$ IQH state on a curved manifold with spacetime metric $g$. The gravitational Chern-Simons term $2\mathrm{CS}_g$ encodes the existence of two Majorana edge modes (a single complex fermion edge mode) on a manifold with boundary. The Laughlin state of holes can be constructed by putting each $f_i$ in a Chern insulator with Chern number $C_i = - 1$. The resulting TQFT is
\begin{equation}\label{eq:JainFCI_U(2)-}
    \begin{aligned}
    &L^{(-)}_{U(2),2/3} = -\frac{1}{4\pi} \Tr \left[(\alpha + A I) d (\alpha + A I) + \frac{2}{3} \alpha^3\right] \\
    &\hspace{0.5cm} - \frac{1}{4\pi} (\Tr \alpha + A) d (\Tr \alpha + A) - 6 \mathrm{CS}_g + \mathrm{CS}[A,g] \\
    &= - \frac{1}{4\pi} \Tr \left[\alpha d \alpha + \frac{2}{3} \alpha^3\right] - \frac{1}{4\pi} \Tr \alpha \, d \Tr \alpha \\
    &\hspace{0.6cm} - \frac{2}{2\pi} \Tr \alpha \,d A - 2 \mathrm{CS}[A,g]  \,. 
    \end{aligned}
\end{equation}
This TQFT can be labeled as $U(2)_{1,6} \times U(1)_{-1}^2$. Following the same steps as in the case of $L^{(+)}_{U(2),2/3}$, we can prove that this state is also topologically equivalent to $\mathrm{Jain}_{2/3}$ (see SM Sec.~\ref{app:JainTQFT}).

Finally, starting from the $U(2)$ gauge theories in \eqref{eq:JainFCI_U(2)+} and \eqref{eq:JainFCI_U(2)-}, we can recover a more familiar $U(1)$ Chern-Simons description of $\mathrm{Jain}_{2/3}$ by modifying the mean-field ansatz for $f_i$. Specifically, if we choose the Chern insulators formed by $f_1, f_2$ to have the same Chern number but distinct band structures, then the $U(2)$ gauge group is Higgsed down to a diagonal $U(1) \times U(1)$ subgroup. Substituting $\alpha = \mathrm{diag}(\alpha_{\uparrow}, \alpha_{\downarrow})$ back to $L^{(+)}_{U(2),2/3}$ in \eqref{eq:JainFCI_U(2)+} and integrating out $\alpha_{\downarrow}$ gives
\begin{comment} 
\begin{equation}
    \begin{aligned}
    L &= - \frac{1}{4\pi} \alpha_{\uparrow} d \alpha_{\uparrow} - \frac{1}{4\pi} \alpha_{\downarrow} d \alpha_{\downarrow} \\
    &\hspace{1cm}- \frac{2}{2\pi} \alpha_{\uparrow} d \alpha_{\downarrow} - \frac{1}{2\pi}(\alpha_{\uparrow} + \alpha_{\downarrow}) d A \,.
    \end{aligned}
\end{equation}
Integrating out $\alpha_{\downarrow}$ then simplifies the Lagrangian to
\end{comment}
\begin{equation}\label{eq:JainFCI_U(1)}
    L_{U(1),2/3} = \frac{3}{4\pi} \alpha_{\uparrow} d \alpha_{\uparrow} + \frac{1}{2\pi} \alpha_{\uparrow} d A + \mathrm{CS}[A,g] \,,
\end{equation}
which is the canonical $U(1)$ CS description of $\mathrm{Jain}_{2/3}$. 

%\ZS{Added a new parton construction for Jain that elucidates the origin of the origin of the boson IQH to trivial transitions} 
Although we arrived at $L^{(\pm)}_{U(2), 2/3}$ through the specific parton construction $c = f_1 f_2 f_3$, this construction is by no means unique. To illustrate this point, we now present an alternative parton construction that also leads to $L^{(\pm)}_{U(2), 2/3}$. In this new construction, we regard $\mathrm{Jain}_{2/3}$ as two copies of the $\nu = 1$ IQH state stacked with another FCI at $\nu = -4/3$. To construct this FCI, we split the electron as $c = \Phi_{\sigma} \epsilon_{\sigma \sigma'} f_{\sigma'}$, introducing a $U(2)$ redundancy such that $\Phi, f$ both carry $SU(2)$ spin $j = 1/2$, but $U(1)$ charge $n = \pm 1$.\footnote{This unusual parton construction dates back to Ref.~\cite{Wen1999_parton} and will be discussed more thoroughly in Ref.~\cite{Shi2026_4eTQC}.} The effective Lagrangian takes the form
\begin{equation}
    L = L_{(1/2,-1)}[\Phi_{\sigma}, \alpha] + L_{(1/2,1)}[f_{\sigma}, \alpha + A I] + 2 \mathrm{CS}[A,g] \,. 
\end{equation}
At $\nu = -4/3$, we can distribute the average Abelian fluxes seen by $\Phi, f$ such that $\nu_{\Phi} = -4$ and $\nu_f = -2$. Putting $\Phi, f$ in bosonic/fermionic IQH states~\cite{Levin2013_bIQH}, we immediately recover $L^{(+)}_{U(2), 2/3}$ upon integrating out $\Phi, f$
\begin{equation}
    \begin{aligned}
    L &= 2 \mathrm{CS}[A,g] + \frac{2}{4\pi} \Tr \left[\alpha d \alpha + \frac{2}{3} \alpha^3\right] - \frac{2}{4\pi} \Tr \alpha \, d \Tr \alpha  \\
    &\hspace{0.5 cm} - \frac{1}{4\pi} \Tr \left[(\alpha + A I) d (\alpha + A I) + \frac{2}{3} \alpha^3\right] - 4 \mathrm{CS}_g \\
    &= \frac{1}{4\pi} \Tr \left[\alpha \, d \alpha + \frac{2}{3} \alpha^3\right] - \frac{2}{4\pi} \Tr \alpha \, d \Tr \alpha - \frac{1}{2\pi} \Tr \alpha \, d A \,. 
    \end{aligned}
\end{equation}
If we redistribute the mean-field fluxes so that $\nu_{\Phi} = -2$ and $\nu_f = -4$, then putting $\Phi, f$ in bosonic/fermionic IQH states gives $L^{(-)}_{U(2),2/3}$
\begin{equation}
    \begin{aligned}
    L &= 2 \mathrm{CS}[A,g] + \frac{1}{4\pi} \Tr \left[\alpha d \alpha + \frac{2}{3} \alpha^3\right] - \frac{1}{4\pi} \Tr \alpha \, d \Tr \alpha  \\
    &- \frac{2}{4\pi} \Tr \left[(\alpha + A I) d (\alpha + A I) + \frac{2}{3} \alpha^3\right] - 8 \mathrm{CS}_g \\
    &= - \frac{1}{4\pi} \Tr \left[\alpha \, d \alpha + \frac{2}{3} \alpha^3\right] - \frac{1}{4\pi} \Tr \alpha \, d \Tr \alpha \\
    &\hspace{0.6cm} - \frac{2}{2\pi} \Tr \alpha \, d A - 2 \mathrm{CS}[A,g] \,.
    \end{aligned}
\end{equation}
As we will see, the existence of multiple parton descriptions for the same Lagrangian provides additional flexibility in constructing Jain-SC transitions.

\begin{comment}
\begin{equation}
    \begin{aligned}
    L[f_i, \alpha + A] &= \sum_{i=1}^2 f^{\dagger}_i \left[D^{\alpha+A}_t - \xi(-i D^{\alpha+A}_i) \right] f_i \,, \\
    D_{\mu} f_i &= \partial_{\mu} f_i - i (\alpha + A I)_{ij} f_j \,,
    \end{aligned}
\end{equation}
\begin{equation}
    \begin{aligned}
    L[f_3, - \Tr \alpha - A] &= f^{\dagger}_3 \left[D^{-\Tr \alpha -A}_t - \xi_3(-i D^{-\Tr \alpha - A}_i) \right] f_3 \,, \\
    D_{\mu} f_3 &= \partial_{\mu} f_3 - i (-\Tr \alpha - A) f_3  \,. 
    \end{aligned}
\end{equation}
\end{comment}

\section{Superconductivity from melting the Jain FCI at \texorpdfstring{$\nu = 2/3$}{}}

The preceding discussion establishes three Lagrangian descriptions (\eqref{eq:JainFCI_U(2)+}, \eqref{eq:JainFCI_U(2)-}, \eqref{eq:JainFCI_U(1)}) of the $\mathrm{Jain}_{2/3}$ TQFT
\begin{equation}
    \mathrm{Jain}_{2/3} = U(2)_{-1,6} = U(2)_{1,6} \times U(1)_{-1}^2 = U(1)_{-3} \times U(1)_1 \,. 
\end{equation}
Armed with this equivalence, we can now describe the phase transitions between $\mathrm{Jain}_{2/3}$ and five distinct superconductors. 

In general, it is useful to classify phase transitions out of fractionalized phases in terms of the quasiparticles that become gapless at the quantum critical point (QCP). In $\mathrm{Jain}_{2/3}$, the two nontrivial anyon types $a_{1/3}/a_{2/3}$ have fractional charge $Q = \frac{1}{3}/Q = \frac{2}{3}$ and energy gaps $\Delta_{1/3}/\Delta_{2/3}$. Phase transitions out of $\mathrm{Jain}_{2/3}$ fall into two broad categories depending on the behavior of anyon energy gaps. In the first category, $\Delta_{2/3}$ vanishes at the QCP while $\Delta_{1/3}$ stays nonzero. Since the electron $c = a_{1/3}^3$ cannot be generated from $a_{2/3}$, the electron stays gapped across the phase transition and the minimum-charge gapless local excitation at the QCP is the charge-2 Cooper pair. We refer to such QCPs as bosonic QCPs, emphasizing the decoupling of the electron from the gapless sector. In the second category, $\Delta_{1/3}$ vanishes at the QCP, thereby closing the energy gaps of all quasiparticles generated by $a_{1/3}$. As a result, both the electron and the Cooper pair become gapless at the QCP, which we refer to as fermionic. We now explore both of these categories.

\subsection{Bosonic QCP: gap closing of \texorpdfstring{$a_{2/3}$}{}}

If $\Delta_{2/3}$ vanishes while $\Delta_{1/3}$ stays finite, then the phase transition occurs in the bosonic sector alone. Within the parton decomposition $c = f_1 f_2 f_3$, $\mathrm{Jain}_{2/3}$ corresponds to the Chern number assignment $C_1 = C_2 = 1, C_3 = -2$. As shown in Ref.~\cite{Shi2024_doping}, $f_1$ and $f_2$ source the anyon $a_{2/3}$ while $f_3$ sources the anyon $a_{1/3}$. Therefore, to construct a bosonic QCP, we must close the energy gaps of $f_1$ and/or $f_2$ while keeping the gap of $f_3$ finite. 

Such a scenario can be naturally realized by tuning the parameters of the microscopic model to induce a band-inversion in the $f_1/f_2$ sector.\footnote{See e.g. Refs.~\cite{Barkeshli2012_LaughlinSF,Lee2018_QED3_FCI,Song2023_QPT_FQAH} for applications of the parton band-inversion mechanism to other transitions out of FCIs.} In the $\mathrm{Jain}_{2/3}$ state, lattice translations $T_x, T_y$ act projectively on $f_1/f_2$ as $T_x T_y T_x^{-1} T_y^{-1} = e^{4\pi i/3}$~\cite{Cheng2016_SET_translation,Song2023_QPT_FQAH}. This projective translation symmetry protects a three-fold degeneracy in the dispersion of $f_1/f_2$. As a result, the Chern numbers of $f_1$ and $f_2$ can only change in integer multiples of 3. Given these symmetry constraints, is there a path from $\mathrm{Jain}_{2/3}$ to a superconductor?

In the simplest $L_{U(1),2/3}$ description, the maximal gauge symmetry $SU(3)$ is broken down to $U(1) \times U(1)$ and the Chern numbers $C_i$ can change independently. By the Ioffe-Larkin composition rule, the physical resistivity tensor $R$ is related to the parton resistivities $R_i$ through $R = \sum_i R_i$~\cite{Ioffe1989_rule}. Since $R_i \propto C_i^{-1}$, superconductivity requires $\sum_i C_i^{-1} = 0$. Holding $C_3 = -2$ fixed, the only way to satisfy $\sum_i C_i^{-1} = 0$ by changing only a single parton Chern number is to choose $C_1 = -2, C_2 = 1$. Since $f_1$ couples to the emergent gauge field $\alpha_{\uparrow} + A$, decreasing $C_1$ by 3 adds a Chern-Simons term for $\alpha_{\uparrow} + A$ with level $= - 3$. The resulting superconductor is therefore\footnote{The superscript $(b)$/$(f)$ in the Lagrangian indicates that the associated Jain-SC transition has only bosonic gapless modes/bosonic + fermionic gapless modes.}
\begin{equation}
    \begin{aligned}
    L^{(b)}_{U(1)} &= L_{U(1),2/3} - \frac{3}{4\pi} (\alpha_{\uparrow} + A) d (\alpha_{\uparrow} + A) - 6 \mathrm{CS}_g \\
    %&\hspace{0.5cm}+ \frac{3}{4\pi} \alpha_{\uparrow} d \alpha_{\uparrow} + \frac{1}{2\pi} \alpha_{\uparrow} d A + \mathrm{CS}[A,g] \\
    &= - \frac{2}{2\pi} \alpha_{\uparrow} d A - 2 \mathrm{CS}[A,g] \,. 
    \end{aligned}
\end{equation}
The final line is precisely the effective field theory for the charge-2e chiral topological SC with $c_- = -2$ on the lower middle part of Fig.~\ref{fig:Jain23_SC}. Since the QCP has three gapless Dirac cones for $f_1$, the critical Lagrangian is precisely the massless $N_f=3$ QED-CS theory
\begin{equation}
    L^{(b)}_{U(1), \rm crit} = i \bar \chi_I \slashed{D}_{\alpha_{\uparrow} + A} \chi_I + L_{U(1), 2/3}[\alpha_{\uparrow}, A] \,.  %\frac{3}{4\pi} \alpha_{\uparrow} d \alpha_{\uparrow} + \frac{1}{2\pi} \alpha_{\uparrow} d A + \mathrm{CS}[A,g] \,.
\end{equation}
previously described in Refs.~\cite{Shi2025_anyon_delocalization,Pichler2025_anyonSC,Wang2025_SCmeltFCI}. 

A more exotic phase transition can be found if we use the non-Abelian $L^{(+)}_{U(2),2/3}$ description for $\mathrm{Jain}_{2/3}$. In the $c = f_1f_2 f_3$ decomposition, the parton mean-field must respect a $U(2)$ gauge symmetry that rotates $f_1$ and $f_2$. As a result, a band-inversion transition must change the Chern numbers $C_1, C_2$ simultaneously by $\pm 3$. Starting from $\{C_i\} = (1, 1, -2)$, the only path to a superconductor is to choose $\Delta C_1 = \Delta C_2 = 3$ so that $\{C_i\}$ goes to $(4,4,-2)$. After integrating out the gapped partons in \eqref{eq:U(2)general}, we find
\begin{equation}\label{eq:Lb_U(2)A}
    \begin{aligned}
    L^{(b)}_{U(2), A} %&= \frac{4}{4\pi} \Tr \left[(\alpha + AI) d (\alpha+ AI) + \frac{2}{3} \alpha^3\right] + 16 \mathrm{CS}_g \\
    %&\hspace{0.5cm} - \frac{2}{4\pi} (\Tr \alpha + A) d (\Tr \alpha + A) - 4 \mathrm{CS}_g \\
    &= \frac{4}{4\pi} \Tr \left[\alpha d \alpha + \frac{2}{3} \alpha^3\right] - \frac{2}{4\pi} \Tr \alpha \, d \Tr \alpha \\
    &\hspace{0.8 cm} + \frac{2}{2\pi} \Tr \alpha \, d A + 6\mathrm{CS}[A,g] \,.
    \end{aligned}
\end{equation}
We recognize this theory as $U(2)_{-4,0} \times U(1)_1^6$. As reviewed in SM Sec.~\ref{app:U(2)SC}, the $U(2)_{-4,0}$ sector describes a charge-4e SC* coexisting with a bosonic topological order $SU(2)_{-4}/\mathbb{Z}_2$, matching the lower left corner of Fig.~\ref{fig:Jain23_SC}. Remarkably, since the elementary $h/4e$ vortex traps a $j = 1/2$ quasiparticle in $SU(2)_{-4}$, gauging the background $A$ transforms the SC* into a non-Abelian topological order. The complete topological data of this gauged theory will be discussed in Ref.~\cite{Shi2026_4eTQC}.

The phase transition from $\mathrm{Jain}_{2/3}$ to $L^{(b)}_{U(2), A}$ involves three pairs of gapless Dirac cones, each transforming in the $(1/2,1)$ representation of $U(2)$. The QCP is therefore an $N_f = 3$ QCD-CS theory with gauge group $U(2)$
\begin{equation}
    L^{(b)}_{U(2), A, \rm crit} = \sum_{I=1}^3 \bar \chi_{I, a} \left(i\slashed{D}_{\alpha + AI}\right)_{ab} \chi_{I,b} + L^{(+)}_{U(2),2/3}[\alpha, A] \,.
\end{equation}

%\ZS{New discussions of a different transition based on $c = \Phi_{\sigma} \epsilon_{\sigma \sigma'} f_{\sigma'}$.} 
Having exhausted all possible Jain-SC transitions driven by band-inversion in the $c = f_1 f_2 f_3$ parton construction, we now turn to the distinct parton construction $c = \Phi_{\sigma} \epsilon_{\sigma \sigma'} f_{\sigma'}$. When $\nu_{\Phi} = -4$ and $\nu_f = -2$, the $a_{2/3}$ anyon is sourced by $f_{\sigma}$. Band-inversion of $f_{\sigma}$ gives the same SC as in \eqref{eq:Lb_U(2)A} and does not generate a new transition pathway. %The projective action of lattice translations implies a three-fold degeneracy in the $f$ spectrum and a band-inversion transition in the $f$ sector changes the Chern number of each $f_{\sigma}$ by $3$. The resulting effective Lagrangian turns out to be equivalent to \eqref{eq:Lb_U(2)A} and we do not get a new superconductor. 

However, when $\nu_{\Phi} = -2$ and $\nu_f = -4$ instead, the $a_{2/3}$ anyon is sourced by $\Phi$. The bosonic analogue of band-inversion is a transition between two invertible bosonic phases across which $\sigma^{\Phi}_{xy}$ jumps by $2$~\cite{Grover2012_bIQHtransition}. Having three degenerate species of $\Phi$ (protected by the projective action of lattice translations) gives $\Delta \sigma^{\Phi}_{xy} = 6$ so that the new state has $\sigma^{\Phi}_{xy} = 4$. Integrating out $\Phi$ and $f$ now gives a distinct superconductor
\begin{equation}
    \begin{aligned}
    L^{(b)}_{U(2), B} &= - \frac{4}{4\pi} \Tr \left[\alpha \, d \alpha + \frac{2}{3} \alpha^3\right] + \frac{2}{4\pi} \Tr \alpha \, d \Tr \alpha \\
    &\hspace{0.6cm} - \frac{2}{2\pi} \Tr \alpha \, d A - 2 \mathrm{CS}[A,g] \,,
    \end{aligned}
\end{equation}
which we recognize as $U(2)_{4,0} \times U(1)_{-1}^2$. As shown in SM Sec.~\ref{app:U(2)SC}, this is a charge-4e SC* with $c_- = 0$ and $SU(2)_{4}/\mathbb{Z}_2$ topological order on the lower right corner of Fig.~\ref{fig:Jain23_SC}. The critical theory involves three species of gapless 2-component bosons $\Phi$, each undergoing a $U(2)$-symmetric boson IQH to insulator transition, which we construct in SM Sec.~\ref{app:BIQH_transition}. Putting all the pieces together, the final critical theory involves $N_f = 3$ emergent two-component fermions $\chi_I$ coupled to a $U(2)$ gauge field $\alpha$ as well as three $U(1)$ gauge fields $a_I$
\begin{equation}
    \begin{aligned}
    L^{(b)}_{U(2), B, \rm crit} &= \sum_{I=1}^3 \left\{ \bar \chi_I (i \slashed{D}_{\alpha - a_I} ) \chi_I + \mathrm{CS}[a_I, g]\right\} \\
    &\hspace{0.8cm} + L^{(-)}_{U(2),2/3}[\alpha,A] \,. 
    \end{aligned}
\end{equation}

\subsection{Fermionic QCP: gap closing of \texorpdfstring{$a_{1/3}$}{}}

We now turn our attention to a more surprising class of phase transitions in which the energy gap of the elementary anyon $a_{1/3}$ vanishes. %The QCPs associated with these phase transitions are genuinely fermionic and have no analog in bosonic FCIs. 
Within the parton decomposition $c = f_1 f_2 f_3$ at filling $\nu = 2/3$, the parton $f_3$ with Chern number $C_3 = -2$ sources the elementary anyon $a_{1/3}$. However, a band-inversion in the $f_3$ sector can only change its Chern number to $C_3 = -5$ or $C_3 = 1$. Both cases lead to a gapped FCI and it appears that a direct transition into a superconductor is impossible.

To overcome this apparent challenge, we switch to a different duality frame $L^{(-)}_{U(2),2/3}$ in which $\mathrm{Jain}_{2/3}$ is realized as an IQH state at $\nu = 1$ stacked with a Laughlin state at $\nu = -1/3$. The same parton decomposition $c = f_1 f_2 f_3$ now assigns Chern numbers $\{C_i\} = (-1, -1, -1)$. Each $f_i$ now sources the elementary Laughlin quasihole with charge $-1/3$ and a band-inversion in any $f_i$ sector closes the energy gap $\Delta_{1/3}$. 

Preserving the $U(2)$ gauge structure, we can contemplate a phase transition in which two of the parton Chern numbers jump by $\pm 3$ simultaneously. The unique Chern numbers (up to permutations) leading to superconductivity are $\{C_i\} = (2, 2, -1)$. Integrating out the gapped partons then gives
\begin{equation}
    \begin{aligned}
        L^{(f)}_{U(2), A} %&= \frac{2}{4\pi} \Tr \left[(\alpha + AI) d (\alpha + AI) + \frac{2}{3} \alpha^3 \right] + 8 \mathrm{CS}_g \\
        %&- \frac{1}{4\pi} (\Tr \alpha + A) d (\Tr \alpha + A) - 2 \mathrm{CS}_g + \mathrm{CS}[A,g] \\
        &= \frac{2}{4\pi} \Tr \left[\alpha d \alpha + \frac{2}{3} \alpha^3 \right] - \frac{1}{4\pi} \Tr \alpha d \Tr \alpha \\
        &\hspace{0.8cm} + \frac{1}{2\pi} \Tr \alpha \, d A + 4\mathrm{CS}[A,g] \,.
    \end{aligned}
\end{equation}
We recognize this theory as $U(2)_{-2,0} \times U(1)_1^4$ (the particle-hole conjugate of this transition has previously appeared in Ref.~\cite{Ma2020_QCDtransition}). Despite the appearance of a non-Abelian CS term, $U(2)_{-2,0}$ secretly describes a charge-2e topological SC with $c_- = -3/2$ and no residual topological order (see SM Sec.~\ref{app:U(2)SC}). Stacking with $U(1)_1^4$ increases the total chiral central charge to $c_- = 5/2$, matching the upper left corner of Fig.~\ref{fig:Jain23_SC}. The critical theory is again $N_f = 3$ QCD-CS, except with a different $U(2)$ Chern-Simons level
\begin{equation}
    L^{(f)}_{U(2), A, \rm crit} = \sum_{I=1}^3 \bar \chi_{I,a} (i\slashed{D}_{\alpha + AI})_{ab} \chi_{I,b} + L^{(-)}_{U(2),2/3}[\alpha, A] \,. 
\end{equation}

Finally, to access the upper right corner of Fig.~\ref{fig:Jain23_SC}, we turn to the exotic parton decomposition $c = \Phi_{\sigma} \epsilon_{\sigma\sigma'} f_{\sigma'}$. When $\nu_{\Phi} = -4, \nu_f = -2$, the $a_{1/3}$ anyon is sourced by the boson $\Phi$. Mimicking the derivation of $L^{(b)}_{U(2), B}$, it is now natural to close the gap $\Delta_{1/3}$ through a transition in the $\Phi$ sector with $\Delta \sigma^{\Phi}_{xy} = 6$. The new state has $\sigma^{\Phi}_{xy} = 2$ and integrating out $\Phi, f$ gives 
\begin{equation}
    \begin{aligned}
        L^{(f)}_{U(2), B} &= - \frac{2}{4\pi} \Tr \left[\alpha d \alpha + \frac{2}{3} \alpha^3\right] \\
        &+ \frac{1}{4\pi} \Tr \alpha \, d \Tr \alpha - \frac{1}{2\pi} \Tr \alpha \, dA \,,
    \end{aligned}
\end{equation}
which we recognize as $U(2)_{2,0}$. Like $L^{(f)}_{U(2),A}$, this new Lagrangian describes a charge-2e topological SC, but with a different chiral central charge $c_- = 3/2$.\footnote{We note that the same superconductor can be obtained by doping $\mathrm{Jain}_{2/3}$, which is equivalent to the particle-hole conjugate of the $k=1$ state in the Read-Rezayi family $\mathrm{RR}_k$~\cite{Shi2025_dopeMR}.} The critical theory again involves three species of Dirac fermions coupled to $\alpha$ and three additional $U(1)$ gauge fields $a_I$
\begin{equation}
    \begin{aligned}
        L^{(f)}_{U(2), B, \rm crit} &= \sum_{I=1}^3 \left\{\bar \chi_I (i\slashed{D}_{\alpha - a_I}) \chi_I + \mathrm{CS}[a_I, g]\right\} \\
        &\hspace{0.5cm} + L^{(+)}_{U(2),2/3}[\alpha, A] \,. 
    \end{aligned}
\end{equation}

\section{Superconductors from doping the FCI}

Let us now explain how each of the SCs constructed so far can also arise from doping the Jain FCIs. While the correspondence is completely general, we will focus on $\mathrm{Jain}_{2/3}$ for concreteness.

Consider the $L^{(+)}_{U(2),2/3}$ representation of the Jain FCI. A complex bosonic field $\Phi$ in the fundamental representation of $U(2)$ sources the anyon with charge $1/3$. Assuming that this is the cheapest anyon excitation, doping induces a finite density of these anyons. Moreover, the dispersion of this anyon has three degenerate minima in the Brillouin zone, as enforced by translation symmetry fractionalization~\cite{Shi2024_doping}. At low density, we therefore model the anyon fluid as three identical species of bosons $\Phi^{(i)}$ coupled to a common $U(2)$ gauge field $\alpha$
\begin{equation}
    L_{\rm doped} = L^{(+)}_{U(2),2/3} + \sum_{i=1}^3 L_{(1/2,1)}[\Phi^{(i)}, \alpha] \,. 
\end{equation}
The equations of motion for the Abelian component $\alpha_0$ impose the constraint $\sum_{i=1}^3 \rho_{\Phi^{(i)}} = \frac{6}{2\pi} \nabla \times \bs{\alpha}_0$. Therefore, each $\Phi^{(i)}$ is at filling $2$ and can form a boson IQH state. Integrating out the gapped $\Phi^{(i)}$ then gives the response theory
\begin{equation}
    \begin{aligned}
    &L_{\rm doped} = L^{(+)}_{U(2),2/3} - \frac{3}{4\pi} \Tr \left[\alpha \, d \alpha + \frac{2}{3} \alpha^3\right] + \frac{3}{4\pi} \Tr \alpha \, d \Tr \alpha \\
    &= - \frac{2}{4\pi} \Tr \left[\alpha \, d \alpha + \frac{2}{3} \alpha^3\right] + \frac{1}{4\pi} \Tr \alpha \, d \Tr \alpha - \frac{1}{2\pi} \Tr \alpha \, d A \,.  
    \end{aligned}
\end{equation}
This final Lagrangian precisely agrees with the SC we found in $L^{(f)}_{U(2),B}$. Following the same logic, one can show that doping the 1/3 anyon in $L^{(-)}_{U(2),2/3}$ gives $L^{(f)}_{U(2),A}$, while doping the 2/3 anyon in $L^{(+)}_{U(2),2/3}$/$L^{(-)}_{U(2),2/3}$ gives the charge-$4e$ SC* in $L^{(b)}_{U(2),A}$/$L^{(b)}_{U(2),B}$.

\section{Discussion}

%\ZS{Wrote a discussion section.}

In summary, leveraging novel Lagrangian descriptions of the Jain FCI at $\nu = 2/3$, we constructed a family of effective field theories that capture continuous phase transitions from $\mathrm{Jain}_{2/3}$ to five distinct superconductors. Besides the simplest topological SC with $c_- = -2$ that has appeared in the literature~\cite{Shi2025_anyon_delocalization,Wang2025_SCmeltFCI,Pichler2025_anyonSC}, the remaining four candidates are non-Abelian topological SCs whose vortices trap non-Abelian zero modes: two of them are charge-2e topological SCs with half-integer $c_-$ and the other two are charge-4e SC*'s with coexistent topological order $SU(2)_{\pm 4}/\mathbb{Z}_2$. In all cases, the Jain-SC phase transitions involve several flavors of gapless Dirac fermions coupled to dynamical $U(1)$ and/or $U(2)$ gauge fields. These transitions are continuous at the mean-field level, but additional numerics are needed to decide whether their continuity survives gauge fluctuations. Given the rapidly expanding list of Chern-Simons matter theories that admit Landau level realizations~\cite{Zhou2024_fuzzy_SpN,Zhou2025_fuzzyCSmatter}, it may be promising to study our theories using the fuzzy sphere regularization~\cite{Zhu2023_fuzzy}. 

In light of our theory, we can revisit the numerical findings of Ref.~\cite{Wang2025_SCmeltFCI}. The most striking observation in Ref.~\cite{Wang2025_SCmeltFCI} is a direct transition from $\mathrm{Jain}_{2/3}$ to a charge-2e chiral topological superconductor with half-integer $c_-$. Two of the phase transitions we discovered--$L^{(f)}_{U(2), A}$ and $L^{(f)}_{U(2), B}$--indeed fit into this general category. However, the value of $c_-$ inferred from the SC pairing symmetry in Ref.~\cite{Wang2025_SCmeltFCI} is $-1/2$, apparently in disagreement with the values $c_- = 5/2$ and $c_- = 3/2$ we found for $L^{(f)}_{U(2), A}$ and $L^{(f)}_{U(2), B}$. Since the connection between pairing symmetry and $c_-$ relies on a weak-coupling BCS analysis, this disagreement is not necessarily physical and should be scrutinized. Ideally, a more direct extraction of $c_-$ within DMRG should resolve this issue definitively. 

A new prediction of our theory is that a continuous transition from $\mathrm{Jain}_{2/3}$ to a charge-2e SC with half-integer $c_-$ must have a vanishing electron energy gap at the quantum critical point. This is in sharp contrast with the transition into an integer-$c_-$ SC previously considered in Ref.~\cite{Shi2025_anyon_delocalization,Wang2025_SCmeltFCI,Pichler2025_anyonSC}, where the Cooper pair gap closes while the electron gap stays open. The closing of the electron gap should lead to a diverging correlation length in the charge-1 sector, which is an important target for future numerical studies. 

Finally, while the effective field theories we proposed capture the competition between FCI and SC, they do not explain why exotic superconductivity wins against an ordinary Fermi liquid at moderate bandwidth. We hope that our constructions can be a useful starting point for addressing this deeper question.

\begin{comment}
Outline:
\begin{enumerate}
    \item Compare with numerics: two of our theories describe transitions from the Jain FCI to a topological charge-2e SC with half-integer chiral central charge. In these cases, the elementary superconducting vortices indeed trap Majorana zero modes. This non-Abelian property matches the numerics by Taige. But the precise value of the chiral central charge is not obviously equivalent. Point out subtlety in numerical extraction of chiral central charge. 
    \item New prediction: the transition between Jain and a topological SC with half-integer chiral central charge must have gapless electron excitations at the QCP. This is in contrast to transitions considered in the past, at which electron gap should remain open. This should be testable in future numerics. 
    \item Mechanism: how does the SC manage to win over the Fermi liquid? Our field theories can be a starting point for answering that question. 
\end{enumerate}
\end{comment}

\begin{acknowledgments}

We would like to thank Ahmed Abouelkomsan, Eduardo Fradkin, Liang Fu, Daniele Guerci and Yin-Chen He for helpful comments. We also thank Zhihuan Dong, Zhaoyu Han, Sri Raghu, Ashvin Vishwanath, Taige Wang and Mike Zaletel for discussions and collaborations on related projects. ZDS was supported by a Leinweber Institute for Theoretical Physics postdoctoral fellowship at Stanford University and in part by Moore grant GBMF8686: Emergent Phenomena in Quantum Systems Theory Center.. TS was supported by NSF grant DMR-2206305. 
\end{acknowledgments}
\bibliography{FQAH-SC}% Produces the bibliography via BibTeX.

@ARTICLE{Park2023_FQAH_TMD,
       author = {{Park}, Heonjoon and {Cai}, Jiaqi and {Anderson}, Eric and {Zhang}, Yinong and {Zhu}, Jiayi and {Liu}, Xiaoyu and {Wang}, Chong and {Holtzmann}, William and {Hu}, Chaowei and {Liu}, Zhaoyu and {Taniguchi}, Takashi and {Watanabe}, Kenji and {Chu}, Jiun-Haw and {Cao}, Ting and {Fu}, Liang and {Yao}, Wang and {Chang}, Cui-Zu and {Cobden}, David and {Xiao}, Di and {Xu}, Xiaodong},
        title = "{Observation of fractionally quantized anomalous Hall effect}",
      journal = {\nat},
         year = 2023,
        month = oct,
       volume = {622},
       number = {7981},
        pages = {74-79},
          doi = {10.1038/s41586-023-06536-0},
archivePrefix = {arXiv},
       eprint = {2308.02657},
 primaryClass = {cond-mat.mes-hall},
       adsurl = {https://ui.adsabs.harvard.edu/abs/2023Natur.622...74P}
}

@ARTICLE{Cai2023_FQAHTMD,
       author = {{Cai}, Jiaqi and {Anderson}, Eric and {Wang}, Chong and {Zhang}, Xiaowei and {Liu}, Xiaoyu and {Holtzmann}, William and {Zhang}, Yinong and {Fan}, Fengren and {Taniguchi}, Takashi and {Watanabe}, Kenji and {Ran}, Ying and {Cao}, Ting and {Fu}, Liang and {Xiao}, Di and {Yao}, Wang and {Xu}, Xiaodong},
        title = "{Signatures of fractional quantum anomalous Hall states in twisted MoTe$_{2}$}",
      journal = {\nat},
         year = 2023,
        month = oct,
       volume = {622},
       number = {7981},
        pages = {63-68},
          doi = {10.1038/s41586-023-06289-w},
archivePrefix = {arXiv},
       eprint = {2304.08470},
 primaryClass = {cond-mat.mes-hall},
       adsurl = {https://ui.adsabs.harvard.edu/abs/2023Natur.622...63C}
}

@ARTICLE{Xu2023_FQAHTMD,
       author = {{Xu}, Fan and {Sun}, Zheng and {Jia}, Tongtong and {Liu}, Chang and {Xu}, Cheng and {Li}, Chushan and {Gu}, Yu and {Watanabe}, Kenji and {Taniguchi}, Takashi and {Tong}, Bingbing and {Jia}, Jinfeng and {Shi}, Zhiwen and {Jiang}, Shengwei and {Zhang}, Yang and {Liu}, Xiaoxue and {Li}, Tingxin},
        title = "{Observation of Integer and Fractional Quantum Anomalous Hall Effects in Twisted Bilayer MoTe$_{2}$}",
      journal = {Physical Review X},
         year = 2023,
        month = jul,
       volume = {13},
       number = {3},
          eid = {031037},
        pages = {031037},
          doi = {10.1103/PhysRevX.13.031037},
archivePrefix = {arXiv},
       eprint = {2308.06177},
 primaryClass = {cond-mat.mes-hall},
       adsurl = {https://ui.adsabs.harvard.edu/abs/2023PhRvX..13c1037X}
}

@ARTICLE{Lu2023_FQAHPenta,
       author = {{Lu}, Zhengguang and {Han}, Tonghang and {Yao}, Yuxuan and {Reddy}, Aidan P. and {Yang}, Jixiang and {Seo}, Junseok and {Watanabe}, Kenji and {Taniguchi}, Takashi and {Fu}, Liang and {Ju}, Long},
        title = "{Fractional quantum anomalous Hall effect in multilayer graphene}",
      journal = {\nat},
         year = 2024,
        month = feb,
       volume = {626},
       number = {8000},
        pages = {759-764},
          doi = {10.1038/s41586-023-07010-7},
archivePrefix = {arXiv},
       eprint = {2309.17436},
 primaryClass = {cond-mat.mes-hall},
       adsurl = {https://ui.adsabs.harvard.edu/abs/2024Natur.626..759L}
}

@ARTICLE{Han2024_chiralSC_penta,
       author = {{Han}, Tonghang and {Lu}, Zhengguang and {Hadjri}, Zach and {Shi}, Lihan and {Wu}, Zhenghan and {Xu}, Wei and {Yao}, Yuxuan and {Cotten}, Armel A. and {Sharifi Sedeh}, Omid and {Weldeyesus}, Henok and {Yang}, Jixiang and {Seo}, Junseok and {Ye}, Shenyong and {Zhou}, Muyang and {Liu}, Haoyang and {Shi}, Gang and {Hua}, Zhenqi and {Watanabe}, Kenji and {Taniguchi}, Takashi and {Xiong}, Peng and {Zumb{\"u}hl}, Dominik M. and {Fu}, Liang and {Ju}, Long},
        title = "{Signatures of chiral superconductivity in rhombohedral graphene}",
      journal = {\nat},
         year = 2025,
        month = jul,
       volume = {643},
       number = {8072},
        pages = {654-661},
          doi = {10.1038/s41586-025-09169-7},
archivePrefix = {arXiv},
       eprint = {2408.15233},
 primaryClass = {cond-mat.mes-hall},
       adsurl = {https://ui.adsabs.harvard.edu/abs/2025Natur.643..654H}
}

@article{Zeng2023_FQAHTMD,
	author = {Zeng, Yihang and Xia, Zhengchao and Kang, Kaifei and Zhu, Jiacheng and Kn{\"u}ppel, Patrick and Vaswani, Chirag and Watanabe, Kenji and Taniguchi, Takashi and Mak, Kin Fai and Shan, Jie},
	da = {2023/10/01},
	doi = {10.1038/s41586-023-06452-3},
	id = {Zeng2023},
	isbn = {1476-4687},
	journal = {Nature},
	number = {7981},
	pages = {69--73},
	title = {Thermodynamic evidence of fractional Chern insulator in moir{\'e}MoTe2},
	ty = {JOUR},
	url = {https://doi.org/10.1038/s41586-023-06452-3},
	volume = {622},
	year = {2023}}

@ARTICLE{Xu2025_SCdopeTMD,
       author = {{Xu}, Fan and {Sun}, Zheng and {Li}, Jiayi and {Zheng}, Ce and {Xu}, Cheng and {Gao}, Jingjing and {Jia}, Tongtong and {Watanabe}, Kenji and {Taniguchi}, Takashi and {Tong}, Bingbing and {Lu}, Li and {Jia}, Jinfeng and {Shi}, Zhiwen and {Jiang}, Shengwei and {Zhang}, Yuanbo and {Zhang}, Yang and {Lei}, Shiming and {Liu}, Xiaoxue and {Li}, Tingxin},
        title = "{Signatures of unconventional superconductivity near reentrant and fractional quantum anomalous Hall insulators}",
      journal = {arXiv e-prints},
         year = 2025,
        month = apr,
          eid = {arXiv:2504.06972},
        pages = {arXiv:2504.06972},
          doi = {10.48550/arXiv.2504.06972},
archivePrefix = {arXiv},
       eprint = {2504.06972},
 primaryClass = {cond-mat.mes-hall},
       adsurl = {https://ui.adsabs.harvard.edu/abs/2025arXiv250406972X}
}

@article{Jain1989_CFframework,
  title = {Composite-fermion approach for the fractional quantum Hall effect},
  author = {Jain, J. K.},
  journal = {Phys. Rev. Lett.},
  volume = {63},
  issue = {2},
  pages = {199--202},
  numpages = {0},
  year = {1989},
  month = {Jul},
  publisher = {American Physical Society},
  doi = {10.1103/PhysRevLett.63.199},
  url = {https://link.aps.org/doi/10.1103/PhysRevLett.63.199}
}

@ARTICLE{Read1999_pair,
       author = {{Read}, N. and {Green}, Dmitry},
        title = "{Paired states of fermions in two dimensions with breaking of parity and time-reversal symmetries and the fractional quantum Hall effect}",
      journal = {\prb},
         year = 2000,
        month = apr,
       volume = {61},
       number = {15},
        pages = {10267-10297},
          doi = {10.1103/PhysRevB.61.10267},
archivePrefix = {arXiv},
       eprint = {cond-mat/9906453},
 primaryClass = {cond-mat.mes-hall},
       adsurl = {https://ui.adsabs.harvard.edu/abs/2000PhRvB..6110267R}
}

@ARTICLE{Ivanov2001_nonabelian,
       author = {{Ivanov}, D.~A.},
        title = "{Non-Abelian Statistics of Half-Quantum Vortices in p-Wave Superconductors}",
      journal = {\prl},
         year = 2001,
        month = jan,
       volume = {86},
       number = {2},
        pages = {268-271},
          doi = {10.1103/PhysRevLett.86.268},
archivePrefix = {arXiv},
       eprint = {cond-mat/0005069},
 primaryClass = {cond-mat.supr-con},
       adsurl = {https://ui.adsabs.harvard.edu/abs/2001PhRvL..86..268I}
}

@article{Sun2011_FCI,
  title = {Nearly Flatbands with Nontrivial Topology},
  author = {Sun, Kai and Gu, Zhengcheng and Katsura, Hosho and Das Sarma, S.},
  journal = {Phys. Rev. Lett.},
  volume = {106},
  issue = {23},
  pages = {236803},
  numpages = {4},
  year = {2011},
  month = {Jun},
  publisher = {American Physical Society},
  doi = {10.1103/PhysRevLett.106.236803},
  url = {https://link.aps.org/doi/10.1103/PhysRevLett.106.236803}
}

@ARTICLE{Sheng2011_FCI,
       author = {{Sheng}, D.~N. and {Gu}, Zheng-Cheng and {Sun}, Kai and {Sheng}, L.},
        title = "{Fractional quantum Hall effect in the absence of Landau levels}",
      journal = {Nature Communications},
         year = 2011,
        month = jul,
       volume = {2},
          eid = {389},
        pages = {389},
          doi = {10.1038/ncomms1380},
archivePrefix = {arXiv},
       eprint = {1102.2658},
 primaryClass = {cond-mat.str-el},
       adsurl = {https://ui.adsabs.harvard.edu/abs/2011NatCo...2..389S}
}

@ARTICLE{Neupert2010_FQAH,
       author = {{Neupert}, Titus and {Santos}, Luiz and {Chamon}, Claudio and {Mudry}, Christopher},
        title = "{Fractional Quantum Hall States at Zero Magnetic Field}",
      journal = {\prl},
         year = 2011,
        month = jun,
       volume = {106},
       number = {23},
          eid = {236804},
        pages = {236804},
          doi = {10.1103/PhysRevLett.106.236804},
archivePrefix = {arXiv},
       eprint = {1012.4723},
 primaryClass = {cond-mat.str-el},
       adsurl = {https://ui.adsabs.harvard.edu/abs/2011PhRvL.106w6804N}
}

@ARTICLE{Tang2011_FQAH,
       author = {{Tang}, Evelyn and {Mei}, Jia-Wei and {Wen}, Xiao-Gang},
        title = "{High-Temperature Fractional Quantum Hall States}",
      journal = {\prl},
         year = 2011,
        month = jun,
       volume = {106},
       number = {23},
          eid = {236802},
        pages = {236802},
          doi = {10.1103/PhysRevLett.106.236802},
archivePrefix = {arXiv},
       eprint = {1012.2930},
 primaryClass = {cond-mat.str-el},
       adsurl = {https://ui.adsabs.harvard.edu/abs/2011PhRvL.106w6802T}
}

@ARTICLE{Regnault2011_FCI,
       author = {{Regnault}, N. and {Bernevig}, B. Andrei},
        title = "{Fractional Chern Insulator}",
      journal = {Physical Review X},
         year = 2011,
        month = oct,
       volume = {1},
       number = {2},
          eid = {021014},
        pages = {021014},
          doi = {10.1103/PhysRevX.1.021014},
archivePrefix = {arXiv},
       eprint = {1105.4867},
 primaryClass = {cond-mat.str-el},
       adsurl = {https://ui.adsabs.harvard.edu/abs/2011PhRvX...1b1014R}
}

@ARTICLE{Song2023_QPT_FQAH,
       author = {{Song}, Xue-Yang and {Zhang}, Ya-Hui and {Senthil}, T.},
        title = "{Phase transitions out of quantum Hall states in moir{\'e} materials}",
      journal = {\prb},
         year = 2024,
        month = feb,
       volume = {109},
       number = {8},
          eid = {085143},
        pages = {085143},
          doi = {10.1103/PhysRevB.109.085143},
archivePrefix = {arXiv},
       eprint = {2308.10903},
 primaryClass = {cond-mat.str-el},
       adsurl = {https://ui.adsabs.harvard.edu/abs/2024PhRvB.109h5143S}
}

@article{Laughlin1988_anyonSC,
  title = {Superconducting Ground State of Noninteracting Particles Obeying Fractional Statistics},
  author = {Laughlin, R. B.},
  journal = {Phys. Rev. Lett.},
  volume = {60},
  issue = {25},
  pages = {2677--2680},
  numpages = {0},
  year = {1988},
  month = {Jun},
  publisher = {American Physical Society},
  doi = {10.1103/PhysRevLett.60.2677},
  url = {https://link.aps.org/doi/10.1103/PhysRevLett.60.2677}
}

@article{Fetter1989_anyonSC_RPA,
  title = {Random-phase approximation in the fractional-statistics gas},
  author = {Fetter, A. L. and Hanna, C. B. and Laughlin, R. B.},
  journal = {Phys. Rev. B},
  volume = {39},
  issue = {13},
  pages = {9679--9681},
  numpages = {0},
  year = {1989},
  month = {May},
  publisher = {American Physical Society},
  doi = {10.1103/PhysRevB.39.9679},
  url = {https://link.aps.org/doi/10.1103/PhysRevB.39.9679}
}

@article{Lee1989_anyonSC,
  title = {Anyon superconductivity and the fractional quantum Hall effect},
  author = {Lee, Dung-Hai and Fisher, Matthew P. A.},
  journal = {Phys. Rev. Lett.},
  volume = {63},
  issue = {8},
  pages = {903--906},
  numpages = {0},
  year = {1989},
  month = {Aug},
  publisher = {American Physical Society},
  doi = {10.1103/PhysRevLett.63.903},
  url = {https://link.aps.org/doi/10.1103/PhysRevLett.63.903}
}

@ARTICLE{Chen1989_anyonSC,
       author = {{Chen}, Yi-Hong and {Wilczek}, Frank and {Witten}, Edward and {Halperin}, Bertrand I.},
        title = "{On Anyon Superconductivity}",
      journal = {International Journal of Modern Physics A},
         year = 1989,
        month = jan,
       volume = {4},
       number = {15},
        pages = {3983},
          doi = {10.1142/S0217751X89001631},
       adsurl = {https://ui.adsabs.harvard.edu/abs/1989IJMPA...4.3983C}
}

@ARTICLE{Levin2013_bIQH,
       author = {{Senthil}, T. and {Levin}, Michael},
        title = "{Integer Quantum Hall Effect for Bosons}",
      journal = {\prl},
         year = 2013,
        month = jan,
       volume = {110},
       number = {4},
          eid = {046801},
        pages = {046801},
          doi = {10.1103/PhysRevLett.110.046801},
archivePrefix = {arXiv},
       eprint = {1206.1604},
 primaryClass = {cond-mat.str-el},
       adsurl = {https://ui.adsabs.harvard.edu/abs/2013PhRvL.110d6801S}
}

@article{Grover2012_bIQHtransition,
  title = {Quantum phase transition between integer quantum Hall states of bosons},
  author = {Grover, Tarun and Vishwanath, Ashvin},
  journal = {Phys. Rev. B},
  volume = {87},
  issue = {4},
  pages = {045129},
  numpages = {10},
  year = {2013},
  month = {Jan},
  publisher = {American Physical Society},
  doi = {10.1103/PhysRevB.87.045129},
  url = {https://link.aps.org/doi/10.1103/PhysRevB.87.045129}
}

@ARTICLE{Witten1989_CSJones,
       author = {{Witten}, Edward},
        title = "{Quantum field theory and the Jones polynomial}",
      journal = {Communications in Mathematical Physics},
         year = 1989,
        month = sep,
       volume = {121},
       number = {3},
        pages = {351-399},
          doi = {10.1007/BF01217730},
       adsurl = {https://ui.adsabs.harvard.edu/abs/1989CMaPh.121..351W}
}

@ARTICLE{Seiberg2016_TQFTgappedbdry,
       author = {{Seiberg}, Nathan and {Witten}, Edward},
        title = "{Gapped boundary phases of topological insulators via weak coupling}",
      journal = {Progress of Theoretical and Experimental Physics},
         year = 2016,
        month = dec,
       volume = {2016},
       number = {12},
          eid = {12C101},
        pages = {12C101},
          doi = {10.1093/ptep/ptw083},
archivePrefix = {arXiv},
       eprint = {1602.04251},
 primaryClass = {cond-mat.str-el},
       adsurl = {https://ui.adsabs.harvard.edu/abs/2016PTEP.2016lC101S}
}

@ARTICLE{Cheng2016_SET_translation,
       author = {{Cheng}, Meng and {Zaletel}, Michael and {Barkeshli}, Maissam and {Vishwanath}, Ashvin and {Bonderson}, Parsa},
        title = "{Translational Symmetry and Microscopic Constraints on Symmetry-Enriched Topological Phases: A View from the Surface}",
      journal = {Physical Review X},
         year = 2016,
        month = dec,
       volume = {6},
       number = {4},
          eid = {041068},
        pages = {041068},
          doi = {10.1103/PhysRevX.6.041068},
archivePrefix = {arXiv},
       eprint = {1511.02263},
 primaryClass = {cond-mat.str-el},
       adsurl = {https://ui.adsabs.harvard.edu/abs/2016PhRvX...6d1068C}
}

@ARTICLE{Cheng2025_orderingqh,
       author = {{Cheng}, Meng and {Musser}, Seth and {Raz}, Amir and {Seiberg}, Nathan and {Senthil}, T.},
        title = "{Ordering the topological order in the fractional quantum Hall effect}",
      journal = {arXiv e-prints},
         year = 2025,
        month = may,
          eid = {arXiv:2505.14767},
        pages = {arXiv:2505.14767},
          doi = {10.48550/arXiv.2505.14767},
archivePrefix = {arXiv},
       eprint = {2505.14767},
 primaryClass = {cond-mat.str-el},
       adsurl = {https://ui.adsabs.harvard.edu/abs/2025arXiv250514767C}
}

@article{kong2014,
  title={Anyon condensation and tensor categories},
  author={Kong, Liang},
  journal={Nuclear Physics B},
  volume={886},
  pages={436--482},
  year={2014},
  publisher={Elsevier},
url={https://www.sciencedirect.com/science/article/pii/S0550321314002223}
}

@article{burnell2018,
  title={Anyon condensation and its applications},
  author={Burnell, Fiona J},
  journal={Annual Review of Condensed Matter Physics},
  volume={9},
  number={1},
  pages={307--327},
  year={2018},
  publisher={Annual Reviews},
url={https://www.annualreviews.org/content/journals/10.1146/annurev-conmatphys-033117-054154}
}

@ARTICLE{Shi2024_doping,
       author = {{Shi}, Zhengyan Darius and {Senthil}, T.},
        title = "{Doping a Fractional Quantum Anomalous Hall Insulator}",
      journal = {Physical Review X},
         year = 2025,
        month = jul,
       volume = {15},
       number = {3},
          eid = {031069},
        pages = {031069},
          doi = {10.1103/kcm5-hx56},
archivePrefix = {arXiv},
       eprint = {2409.20567},
 primaryClass = {cond-mat.str-el},
       adsurl = {https://ui.adsabs.harvard.edu/abs/2025PhRvX..15c1069S}
}

@ARTICLE{Kim2024_anyonSC,
       author = {{Kim}, Minho and {Timmel}, Abigail and {Ju}, Long and {Wen}, Xiao-Gang},
        title = "{Topological chiral superconductivity beyond pairing in a Fermi liquid}",
      journal = {\prb},
         year = 2025,
        month = jan,
       volume = {111},
       number = {1},
          eid = {014508},
        pages = {014508},
          doi = {10.1103/PhysRevB.111.014508},
archivePrefix = {arXiv},
       eprint = {2409.18067},
 primaryClass = {cond-mat.str-el},
       adsurl = {https://ui.adsabs.harvard.edu/abs/2025PhRvB.111a4508K}
}

@ARTICLE{Divic2024_HofHubb,
       author = {{Divic}, Stefan and {Cr{\'e}pel}, Valentin and {Soejima}, Tomohiro and {Song}, Xue-Yang and {Millis}, Andrew J. and {Zaletel}, Michael P. and {Vishwanath}, Ashvin},
        title = "{Anyon superconductivity from topological criticality in a Hofstadter-Hubbard model}",
      journal = {Proceedings of the National Academy of Science},
         year = 2025,
        month = aug,
       volume = {122},
       number = {33},
          eid = {e2426680122},
        pages = {e2426680122},
          doi = {10.1073/pnas.2426680122},
archivePrefix = {arXiv},
       eprint = {2410.18175},
 primaryClass = {cond-mat.str-el},
       adsurl = {https://ui.adsabs.harvard.edu/abs/2025PNAS..12226680D}
}

@ARTICLE{Shi2025_dopeMR,
       author = {{Darius Shi}, Zhengyan and {Zhang}, Carolyn and {Senthil}, T.},
        title = "{Doping lattice non-abelian quantum Hall states}",
      journal = {arXiv e-prints},
         year = 2025,
        month = may,
          eid = {arXiv:2505.02893},
        pages = {arXiv:2505.02893},
          doi = {10.48550/arXiv.2505.02893},
archivePrefix = {arXiv},
       eprint = {2505.02893},
 primaryClass = {cond-mat.str-el},
       adsurl = {https://ui.adsabs.harvard.edu/abs/2025arXiv250502893D}
}

@ARTICLE{Shi2025_anyon_delocalization,
       author = {{Darius Shi}, Zhengyan and {Senthil}, T.},
        title = "{Anyon delocalization transitions out of a disordered FQAH insulator}",
      journal = {arXiv e-prints},
         year = 2025,
        month = jun,
          eid = {arXiv:2506.02128},
        pages = {arXiv:2506.02128},
          doi = {10.48550/arXiv.2506.02128},
archivePrefix = {arXiv},
       eprint = {2506.02128},
 primaryClass = {cond-mat.str-el},
       adsurl = {https://ui.adsabs.harvard.edu/abs/2025arXiv250602128D}
}

@ARTICLE{Nosov2025_plateau,
       author = {{Nosov}, Pavel A. and {Han}, Zhaoyu and {Khalaf}, Eslam},
        title = "{Anyon superconductivity and plateau transitions in doped fractional quantum anomalous Hall insulators}",
      journal = {arXiv e-prints},
         year = 2025,
        month = jun,
          eid = {arXiv:2506.02108},
        pages = {arXiv:2506.02108},
          doi = {10.48550/arXiv.2506.02108},
archivePrefix = {arXiv},
       eprint = {2506.02108},
 primaryClass = {cond-mat.str-el},
       adsurl = {https://ui.adsabs.harvard.edu/abs/2025arXiv250602108N}
}

@ARTICLE{Pichler2025_anyonSC,
       author = {{Pichler}, Fabian and {Kuhlenkamp}, Clemens and {Knap}, Michael and {Vishwanath}, Ashvin},
        title = "{Microscopic Mechanism of Anyon Superconductivity Emerging from Fractional Chern Insulators}",
      journal = {arXiv e-prints},
         year = 2025,
        month = jun,
          eid = {arXiv:2506.08000},
        pages = {arXiv:2506.08000},
          doi = {10.48550/arXiv.2506.08000},
archivePrefix = {arXiv},
       eprint = {2506.08000},
 primaryClass = {cond-mat.str-el},
       adsurl = {https://ui.adsabs.harvard.edu/abs/2025arXiv250608000P}
}

@ARTICLE{Han2025_anyonexciton,
       author = {{Han}, Zhaoyu and {Wang}, Taige and {Dong}, Zhihuan and {Zaletel}, Michael P. and {Vishwanath}, Ashvin},
        title = "{Anyon superfluidity of excitons in quantum Hall bilayers}",
      journal = {arXiv e-prints},
         year = 2025,
        month = aug,
          eid = {arXiv:2508.14894},
        pages = {arXiv:2508.14894},
          doi = {10.48550/arXiv.2508.14894},
archivePrefix = {arXiv},
       eprint = {2508.14894},
 primaryClass = {cond-mat.str-el},
       adsurl = {https://ui.adsabs.harvard.edu/abs/2025arXiv250814894H}
}

@ARTICLE{Nakajima2025_thermo_anyon,
       author = {{Nakajima}, Yuto and {Mehta}, Umang and {Goldman}, Hart},
        title = "{Thermodynamics of dilute anyon gases from fusion constraints}",
      journal = {arXiv e-prints},
         year = 2025,
        month = aug,
          eid = {arXiv:2508.14961},
        pages = {arXiv:2508.14961},
          doi = {10.48550/arXiv.2508.14961},
archivePrefix = {arXiv},
       eprint = {2508.14961},
 primaryClass = {cond-mat.str-el},
       adsurl = {https://ui.adsabs.harvard.edu/abs/2025arXiv250814961N}
}

@ARTICLE{Kuhlenkamp2025_HofHubb,
       author = {{Kuhlenkamp}, Clemens and {Divic}, Stefan and {Zaletel}, Michael P. and {Soejima}, Tomohiro and {Vishwanath}, Ashvin},
        title = "{Robust superconductivity upon doping chiral spin liquid and Chern insulators in a Hubbard-Hofstadter model}",
      journal = {arXiv e-prints},
         year = 2025,
        month = sep,
          eid = {arXiv:2509.02675},
        pages = {arXiv:2509.02675},
          doi = {10.48550/arXiv.2509.02675},
archivePrefix = {arXiv},
       eprint = {2509.02675},
 primaryClass = {cond-mat.str-el},
       adsurl = {https://ui.adsabs.harvard.edu/abs/2025arXiv250902675K}
}

@article{Barkeshli2012_LaughlinSF,
  title = {Continuous transition between fractional quantum Hall and superfluid states},
  author = {Barkeshli, Maissam and McGreevy, John},
  journal = {Phys. Rev. B},
  volume = {89},
  issue = {23},
  pages = {235116},
  numpages = {6},
  year = {2014},
  month = {Jun},
  publisher = {American Physical Society},
  doi = {10.1103/PhysRevB.89.235116},
  url = {https://link.aps.org/doi/10.1103/PhysRevB.89.235116}
}

@ARTICLE{Wang2025_LaughlinSF,
       author = {{Wang}, Taige and {Song}, Xue-Yang and {Zaletel}, Michael P. and {Senthil}, T.},
        title = "{Emergent QED$_3$ at the bosonic Laughlin state to superfluid transition}",
      journal = {arXiv e-prints},
         year = 2025,
        month = jul,
          eid = {arXiv:2507.07611},
        pages = {arXiv:2507.07611},
          doi = {10.48550/arXiv.2507.07611},
archivePrefix = {arXiv},
       eprint = {2507.07611},
 primaryClass = {cond-mat.str-el},
       adsurl = {https://ui.adsabs.harvard.edu/abs/2025arXiv250707611W}
}

@ARTICLE{Wang2025_SCmeltFCI,
       author = {{Wang}, Taige and {Zaletel}, Michael P.},
        title = "{Chiral superconductivity near a fractional Chern insulator}",
      journal = {arXiv e-prints},
         year = 2025,
        month = jul,
          eid = {arXiv:2507.07921},
        pages = {arXiv:2507.07921},
          doi = {10.48550/arXiv.2507.07921},
archivePrefix = {arXiv},
       eprint = {2507.07921},
 primaryClass = {cond-mat.str-el},
       adsurl = {https://ui.adsabs.harvard.edu/abs/2025arXiv250707921W}
}

@ARTICLE{Goncalves2025_anyondisp,
       author = {{Gon{\c{c}}alves}, Miguel and {Mendez-Valderrama}, Juan Felipe and {Herzog-Arbeitman}, Jonah and {Yu}, Jiabin and {Xu}, Xiaodong and {Xiao}, Di and {Bernevig}, B. Andrei and {Regnault}, Nicolas},
        title = "{Spinless and spinful charge excitations in moir{\'e} Fractional Chern Insulators}",
      journal = {arXiv e-prints},
         year = 2025,
        month = jun,
          eid = {arXiv:2506.05330},
        pages = {arXiv:2506.05330},
          doi = {10.48550/arXiv.2506.05330},
archivePrefix = {arXiv},
       eprint = {2506.05330},
 primaryClass = {cond-mat.str-el},
       adsurl = {https://ui.adsabs.harvard.edu/abs/2025arXiv250605330G}
}

@ARTICLE{Schleith2025_anyondisp,
       author = {{Schleith}, Mina-Lou and {Soejima}, Tomohiro and {Khalaf}, Eslam},
        title = "{Anyon dispersion from non-uniform magnetic field on the sphere}",
      journal = {arXiv e-prints},
         year = 2025,
        month = jun,
          eid = {arXiv:2506.11211},
        pages = {arXiv:2506.11211},
          doi = {10.48550/arXiv.2506.11211},
archivePrefix = {arXiv},
       eprint = {2506.11211},
 primaryClass = {cond-mat.str-el},
       adsurl = {https://ui.adsabs.harvard.edu/abs/2025arXiv250611211S}
}

@ARTICLE{Yan2025_anyondisp,
       author = {{Yan}, Zihan and {Li}, Qingchen and {Soejima}, Tomohiro and {Khalaf}, Eslam},
        title = "{Anyon Dispersion in Aharonov-Casher Bands and Implications for Twisted MoTe${}_2$}",
      journal = {arXiv e-prints},
         year = 2025,
        month = dec,
          eid = {arXiv:2512.15863},
        pages = {arXiv:2512.15863},
          doi = {10.48550/arXiv.2512.15863},
archivePrefix = {arXiv},
       eprint = {2512.15863},
 primaryClass = {cond-mat.str-el},
       adsurl = {https://ui.adsabs.harvard.edu/abs/2025arXiv251215863Y}
}

@article{Ioffe1989_rule,
  title = {Gapless fermions and gauge fields in dielectrics},
  author = {Ioffe, L. B. and Larkin, A. I.},
  journal = {Phys. Rev. B},
  volume = {39},
  issue = {13},
  pages = {8988--8999},
  numpages = {0},
  year = {1989},
  month = {May},
  publisher = {American Physical Society},
  doi = {10.1103/PhysRevB.39.8988},
  url = {https://link.aps.org/doi/10.1103/PhysRevB.39.8988}
}

@ARTICLE{Lee2018_QED3_FCI,
       author = {{Lee}, Jong Yeon and {Wang}, Chong and {Zaletel}, Michael P. and {Vishwanath}, Ashvin and {He}, Yin-Chen},
        title = "{Emergent Multi-Flavor QED$_{3}$ at the Plateau Transition between Fractional Chern Insulators: Applications to Graphene Heterostructures}",
      journal = {Physical Review X},
         year = 2018,
        month = jul,
       volume = {8},
       number = {3},
          eid = {031015},
        pages = {031015},
          doi = {10.1103/PhysRevX.8.031015},
archivePrefix = {arXiv},
       eprint = {1802.09538},
 primaryClass = {cond-mat.str-el},
       adsurl = {https://ui.adsabs.harvard.edu/abs/2018PhRvX...8c1015L}
}

@article{Kol1993_Jainlattice,
  title = {Fractional quantum Hall effect in a periodic potential},
  author = {Kol, A. and Read, N.},
  journal = {Phys. Rev. B},
  volume = {48},
  issue = {12},
  pages = {8890--8898},
  numpages = {0},
  year = {1993},
  month = {Sep},
  publisher = {American Physical Society},
  doi = {10.1103/PhysRevB.48.8890},
  url = {https://link.aps.org/doi/10.1103/PhysRevB.48.8890}
}

@ARTICLE{Shi2026_4eTQC,
       author = {{Darius Shi}, Zhengyan and {Han}, Zhaoyu and {Raghu}, Srinivas and {Vishwanath}, Ashvin},
        title = "{Charge-$4e$ superconductor with parafermionic vortices: A path to universal topological quantum computation}",
      journal = {arXiv e-prints},
         year = 2026,
        month = feb,
          eid = {arXiv:2602.06963},
        pages = {arXiv:2602.06963},
          doi = {10.48550/arXiv.2602.06963},
archivePrefix = {arXiv},
       eprint = {2602.06963},
 primaryClass = {cond-mat.str-el},
       adsurl = {https://ui.adsabs.harvard.edu/abs/2026arXiv260206963D}
}

@ARTICLE{Zhu2023_fuzzy,
       author = {{Zhu}, Wei and {Han}, Chao and {Huffman}, Emilie and {Hofmann}, Johannes S. and {He}, Yin-Chen},
        title = "{Uncovering Conformal Symmetry in the 3D Ising Transition: State-Operator Correspondence from a Quantum Fuzzy Sphere Regularization}",
      journal = {Physical Review X},
         year = 2023,
        month = apr,
       volume = {13},
       number = {2},
          eid = {021009},
        pages = {021009},
          doi = {10.1103/PhysRevX.13.021009},
archivePrefix = {arXiv},
       eprint = {2210.13482},
 primaryClass = {cond-mat.stat-mech},
       adsurl = {https://ui.adsabs.harvard.edu/abs/2023PhRvX..13b1009Z}
}

@ARTICLE{Zhou2024_fuzzy_SpN,
       author = {{Zhou}, Zheng and {He}, Yin-Chen},
        title = "{3D Conformal Field Theories with Sp(N) Global Symmetry on a Fuzzy Sphere}",
      journal = {\prl},
         year = 2025,
        month = jul,
       volume = {135},
       number = {2},
          eid = {026504},
        pages = {026504},
          doi = {10.1103/xstj-xvcy},
archivePrefix = {arXiv},
       eprint = {2410.00087},
 primaryClass = {hep-th},
       adsurl = {https://ui.adsabs.harvard.edu/abs/2025PhRvL.135b6504Z}
}

@ARTICLE{Zhou2025_fuzzyCSmatter,
       author = {{Zhou}, Zheng and {Wang}, Chong and {He}, Yin-Chen},
        title = "{Chern-Simons-matter conformal field theory on fuzzy sphere: Confinement transition of Kalmeyer-Laughlin chiral spin liquid}",
      journal = {arXiv e-prints},
         year = 2025,
        month = jul,
          eid = {arXiv:2507.19580},
        pages = {arXiv:2507.19580},
          doi = {10.48550/arXiv.2507.19580},
archivePrefix = {arXiv},
       eprint = {2507.19580},
 primaryClass = {cond-mat.str-el},
       adsurl = {https://ui.adsabs.harvard.edu/abs/2025arXiv250719580Z}
}

@ARTICLE{Wen1999_parton,
       author = {{Wen}, Xiao-Gang},
        title = "{Projective construction of non-Abelian quantum Hall liquids}",
      journal = {\prb},
         year = 1999,
        month = sep,
       volume = {60},
       number = {12},
        pages = {8827-8838},
          doi = {10.1103/PhysRevB.60.8827},
archivePrefix = {arXiv},
       eprint = {cond-mat/9811111},
 primaryClass = {cond-mat.mes-hall},
       adsurl = {https://ui.adsabs.harvard.edu/abs/1999PhRvB..60.8827W}
}

@ARTICLE{Ma2020_QCDtransition,
       author = {{Ma}, Ruochen and {He}, Yin-Chen},
        title = "{Emergent QCD$_{3}$ quantum phase transitions of fractional Chern insulators}",
      journal = {Physical Review Research},
         year = 2020,
        month = sep,
       volume = {2},
       number = {3},
          eid = {033348},
        pages = {033348},
          doi = {10.1103/PhysRevResearch.2.033348},
archivePrefix = {arXiv},
       eprint = {2003.05954},
 primaryClass = {cond-mat.str-el},
       adsurl = {https://ui.adsabs.harvard.edu/abs/2020PhRvR...2c3348M}
}

@ARTICLE{Sohal2017_fluxattach_FCI,
       author = {{Sohal}, Ramanjit and {Santos}, Luiz H. and {Fradkin}, Eduardo},
        title = "{Chern-Simons composite fermion theory of fractional Chern insulators}",
      journal = {\prb},
         year = 2018,
        month = mar,
       volume = {97},
       number = {12},
          eid = {125131},
        pages = {125131},
          doi = {10.1103/PhysRevB.97.125131},
archivePrefix = {arXiv},
       eprint = {1707.06118},
 primaryClass = {cond-mat.str-el},
       adsurl = {https://ui.adsabs.harvard.edu/abs/2018PhRvB..97l5131S}
}

@ARTICLE{Guerci2025_FCISC,
       author = {{Guerci}, Daniele and {Abouelkomsan}, Ahmed and {Fu}, Liang},
        title = "{From Fractionalization to Chiral Topological Superconductivity in a Flat Chern Band}",
      journal = {\prl},
         year = 2025,
        month = oct,
       volume = {135},
       number = {18},
          eid = {186601},
        pages = {186601},
          doi = {10.1103/zm39-dstj},
archivePrefix = {arXiv},
       eprint = {2506.10938},
 primaryClass = {cond-mat.supr-con},
       adsurl = {https://ui.adsabs.harvard.edu/abs/2025PhRvL.135r6601G}
}

@ARTICLE{Guerci2026_TSC_vorlat,
       author = {{Guerci}, Daniele and {Abouelkomsan}, Ahmed and {Fu}, Liang},
        title = "{Topological superconductivity with emergent vortex lattice in twisted semiconductors}",
      journal = {arXiv e-prints},
         year = 2026,
        month = feb,
          eid = {arXiv:2602.15106},
        pages = {arXiv:2602.15106},
          doi = {10.48550/arXiv.2602.15106},
archivePrefix = {arXiv},
       eprint = {2602.15106},
 primaryClass = {cond-mat.supr-con},
       adsurl = {https://ui.adsabs.harvard.edu/abs/2026arXiv260215106G}
}

\onecolumngrid
\appendix

\newpage 

\section{\texorpdfstring{$U(2)$}{} gauge theory for fermionic Jain states}\label{app:JainTQFT}

In this section, we construct an unusual $U(2)$ gauge theory representation of the fermionic Jain FCI at lattice filling $\nu = p/(2p+1)$, henceforth denoted $\mathrm{Jain}_p$. Taking $p = -2$ recovers the $\nu = 2/3$ Jain FCI (for which we used a different notation $\mathrm{Jain}_{2/3}$ in the main text). Given that the Jain topological order is Abelian for all $p$, this non-Abelian gauge theory description may appear to be unnecessarily complicated. However, as we have seen in the main text, the $U(2)$ structure is indispensible for capturing several distinct continuous phase transitions into neighboring exotic superconductors. 

Following the main text, we consider a microscopic translation-invariant lattice model of electrons at charge filling $\nu = p/(2p+1)$, with a background $2 \pi \mathbb{Z}$ magnetic flux per unit cell. We decompose the electron operator $c(\bs{r})$ as $c = f_1 f_2 f_3$, where $f_1, f_2, f_3$ are fermionic partons. This decomposition has a local $SU(3)$ gauge redundancy $f_i \rightarrow U_{ij} f_j$. In what follows, we choose a mean-field ansatz that Higgses the $SU(3)$ gauge group down to a $U(2)$ subgroup that acts on the fermions as 
\begin{equation}
    f_{i=1,2} \rightarrow \sum_{j=1}^2 \tilde U_{ij} f_j \,, \quad f_3 \rightarrow \det \tilde U^{-1} f_3 \,, \quad \tilde U \in U(2) \,.  
\end{equation}
We assign a global electric charge $+1$ to $f_1, f_2$ and $-1$ to $f_3$ so that the physical electron carries the correct electric charge $+1$. The action of $U(2)$ on $f_1, f_2, f_3$ implies that $(f_1, f_2)$ transform in the $(1/2,1)$ representation of $U(2)$ while $f_3$ transforms in the $(0,-2)$ representation of $U(2)$. The low-energy effective Lagrangian therefore takes the general form 
\begin{equation}
    L = L_{(1/2,1)}[f_i, \alpha + AI] + L_{(0,-2)}[\psi, -\Tr \alpha - A] \,, 
\end{equation}
where $\alpha$ is a dynamical $U(2)$ gauge field, $A$ is the external electromagnetic gauge field (technically a $\mathrm{spin}_{\mathbb{C}}$ connection), and $L_{(j,n)}$ describes a matter field in the $(j,n)$ representation minimally coupled to $\alpha$. 

Following the standard construction of Jain states, we can choose a flux assignment for $\Tr \alpha$ such that $f_1, f_2, f_3$ form Chern insulators with Chern numbers $C_1 = C_2 = 1$ and $C_3 = p$. After integrating out the gapped partons, we obtain a TQFT
\begin{equation}
    \begin{aligned}
    L^{(+)}_{U(2),p} &= \frac{1}{4\pi} \Tr \left[(\alpha + AI) d (\alpha + AI)+ \frac{2}{3} \alpha^3\right] + \frac{p}{4\pi} (\Tr \alpha + A) \, d (\Tr \alpha + A) + (2p+4) \mathrm{CS}_g \\
    &= \frac{1}{4\pi} \Tr \left[\alpha d \alpha + \frac{2}{3} \alpha^3\right] +  \frac{p}{4\pi} \Tr \alpha \, d \Tr \alpha + \frac{p+1}{2\pi} \Tr \alpha \, d A + (p+2) \mathrm{CS}[A,g] \,.
    \end{aligned}
\end{equation}
In this Lagrangian, $g$ is a background spacetime metric and $\mathrm{CS}_g$ is a gravitational Chern-Simons term, which encodes the existence of a single Majorana edge mode when the system is placed on a manifold with boundary. The term $\mathrm{CS}[A,g]$ is a short hand for the topological response of the $\nu = 1$ IQH state $\mathrm{CS}[A,g] = 1/(4\pi) A dA + 2 \mathrm{CS}_g$. This expression simultaneously captures the quantized Hall conductance $\sigma_{xy} = 1$ and the existence of two Majorana edge modes which combine into the standard complex fermion edge mode of the IQH state. By decomposing $\alpha$ into its $U(1)$ and $SU(2)$ components, we find that the $U(1)$ Chern-Simons level is $-2-4p$ while the $U(2)$ Chern-Simons level is $-1$. Following the standard convention, we therefore refer to $L^{(+)}_{U(2),p}$ as $U(2)_{-1,-2-4p} \times U(1)_1^{p+2}$. Since the mean-field wavefunction is identical to the Jain state wavefunction, we know that this complicated TQFT secretly describes the Abelian topological order $\mathrm{Jain}_p$. 

The same conclusion can be reached through a more formal route. From Ref.~\cite{Cheng2025_orderingqh}, we know that $\mathrm{Jain}_p$ is the unique topological order at $\nu = p/(2p+1)$ with $|2p+1|$-fold torus ground state degeneracy, Hall conductance $\sigma_{xy} = p/(2p+1)$, and chiral central charge $c_- = p+1 - \sgn(p)$. Let us now verify that all of these properties are satisfied by $U(2)_{-1,-2-4p} \times U(1)_1^{p+2}$. 

As a topological order, $U(1)_1^{p+2}$ is an invertible theory and does not contribute any torus ground state degeneracy. Therefore, the ground degeneracy is entire determined by the $U(2)_{-1, -2-4p}$ sector, which can be decomposed as
\begin{equation}
    U(2)_{-1, -2-4p} = \frac{SU(2)_{-1} \times U(1)_{-2-4p}}{\mathbb{Z}_2} \,. 
\end{equation}
In the field-theoretic language, the $\mathbb{Z}_2$ quotient corresponds to gauging a fermionic $\mathbb{Z}_2$ 1-form symmetry generated by the anyon in $SU(2)_{-1} \times U(1)_{-2-4p}$ labeled by the $SU(2)$ spin $j = 1/2$ and the $U(1)$ charge $n = - (2p+1)$. Following the general dictionary of  Refs.~\cite{kong2014,burnell2018}, we immediately deduce the squared quantum dimension
\begin{equation}
    \mathcal{D}^2[U(2)_{-1, -2-4p}] = \frac{\mathcal{D}^2[SU(2)_{-1}] \times \mathcal{D}^2[U(1)_{-2-4p}]}{2} = \frac{2 \times 2|2p+1|}{2} = 2 |2p+1| \,. 
\end{equation}
The torus ground state degeneracy of a fermionic topological order with squared quantum dimension $\mathcal{D}^2$ is precisely $\mathcal{D}^2/2$. Therefore, $U(2)_{-1,-2-4p} \times U(1)_1^{p+2}$ has a torus ground state degeneracy $|2p+1|$, in agreement with $\mathrm{Jain}_p$. 

As for the Hall conductance, since $\alpha$ only couples to $A$ through its Abelian component $\Tr \alpha$, we can calculate the Hall conductance by restricting to gauge field configurations with $\alpha = \alpha_0 I$. This leads to an Abelianized Lagrangian
\begin{equation}
    L^{(+), \rm Abelianized}_{U(2),p} \rightarrow \frac{4p+2}{4\pi} \alpha_0 d \alpha_0 + \frac{2(p+1)}{2\pi} \alpha_0 d A + (p+2) \mathrm{CS}[A,g] \,. 
\end{equation}
From this formula, we can read off the Hall conductance which matches the corresponding value for $\mathrm{Jain}_p$
\begin{equation}
    \sigma_{xy} = \frac{4 (p+1)^2}{-(4p+2)} + (p+2) = \frac{p}{2p+1} \,. 
\end{equation}

Finally, for the chiral central charge $c_-$, we recall that $SU(2)_{-1}$ has $c_- = -1$ and $U(1)_{-2-4p}$ has $c_- = \sgn(p)$. Since discrete gauging does not modify $c_-$, the quotient theory $U(2)_{-1,-2-4p}$ therefore has $c_- = -1 + \sgn(p)$. Stacking with the $(p+2)$ copies of IQH state gives a total chiral central charge $c_- = p+1 + \sgn(p)$, which again matches the value for $\mathrm{Jain}_p$. Taken together, these three checks establish that $U(2)_{-1,-2-4p} \times U(1)_1^{p+2}$ is topologically equivalent to $\mathrm{Jain}_p$. It is also easy to see that $f_1, f_2$ source the vison with charge $p/(2p+1)$, while $\psi$ sources the minimally charged anyon with charge $1/(2p+1)$. 

An alternative $U(2)$ description of the same state can be obtained by viewing the Jain state at $\nu = p/(2p+1)$ as the particle-hole conjugate of the Jain state at filling $\tilde \nu = \frac{\tilde p}{2 \tilde p + 1}$ with $\tilde p = - (p+1)$. In terms of parton construction, this corresponds to stacking an IQH state with additional electrons $c$ at filling $\nu = - \frac{\tilde p}{2 \tilde p + 1}$ and fractionalizing $c$ as $c = f_1 f_2 f_3$. Putting $f_i$ into Chern insulators with Chern numbers $C_1 = C_2 = -1, C_3 = \tilde p = - \tilde p$. After integrating out the gapped partons, we obtain the alternative description 
\begin{equation}
    \begin{aligned}
    L^{(-)}_{U(2), p} &= \mathrm{CS}[A,g] - L^{(+)}_{U(2),-(p+1)} \\
    &= - \frac{1}{4\pi} \Tr \left[(\alpha + AI) d (\alpha + AI) + \frac{2}{3} \alpha^3\right] +\frac{p+1}{4\pi} (\Tr \alpha + A) \, d (\Tr \alpha + A) + (2p-4) \mathrm{CS}_g + \mathrm{CS}[A,g] \\
    &= - \frac{1}{4\pi} \Tr \left[\alpha d \alpha + \frac{2}{3} \alpha^3\right] + \frac{p+1}{4\pi} \Tr \alpha \, d \Tr \alpha + \frac{p}{2\pi} \Tr \alpha \, d A + p \mathrm{CS}[A,g] \,. 
    \end{aligned}
\end{equation}
This alternative TQFT describes the $U(2)_{1,-4p-2} \times U(1)_1^p$ topological order. One can check that the chiral central charge is $c_- = p + 1 - \sgn(p)$ and the Hall conductance is $\sigma_{xy} = p - p^2/2(2p+1) = p/(2p+1)$, again in agreement with $\mathrm{Jain}_p$.\footnote{Setting $p = 1$ recovers the $\mathrm{RR}_1$ description of the Laughlin state at $\nu = 1/3$, as first pointed out in Ref.~\cite{Shi2025_dopeMR}.} 

Finally, if we choose a parton mean-field in which $f_1, f_2$ have the same Chern number but different band structures, the $U(2)$ gauge symmetry is Higgsed down to $U(1) \times U(1)$. Plugging $\alpha = \mathrm{diag}(\alpha_{\uparrow}, \alpha_{\downarrow})$ back into $L^{(-)}_{U(2),p}$ gives 
\begin{equation}
    L_{U(1) \times U(1), p} = \frac{1}{4\pi} (\alpha_{\uparrow} d \alpha_{\uparrow} + \alpha_{\downarrow} d \alpha_{\downarrow}) + \frac{p}{4\pi} (\alpha_{\uparrow} + \alpha_{\downarrow}) d (\alpha_{\uparrow} + \alpha_{\downarrow}) + \frac{p+1}{2\pi} (\alpha_{\uparrow} + \alpha_{\downarrow}) d A + (p+2) \mathrm{CS}[A,g] \,. 
\end{equation}
Making a shift $\alpha_{\uparrow} \rightarrow \alpha_{\uparrow} - \alpha_{\downarrow}$, we obtain a simpler Lagrangian
\begin{equation}
    L_{U(1) \times U(1), p} = \frac{p+1}{4\pi} \alpha_{\uparrow} d \alpha_{\uparrow} + \frac{2}{2\pi} \alpha_{\downarrow} d \alpha_{\downarrow} - \frac{1}{2\pi} \alpha_{\uparrow} d \alpha_{\downarrow} + \frac{p+1}{2\pi} \alpha_{\uparrow} d A + (p+2) \mathrm{CS}[A,g] \,. 
\end{equation}
We recognize this K-matrix theory as the canonical description of $\mathrm{Jain}_p$ (see e.g. Ref.~\cite{Shi2024_doping} for a review). 

In summary, in addition to the standard K-matrix description of $\mathrm{Jain}_p$, we obtained two additional non-Abelian descriptions $U(2)_{1, -4p-2} \times U(1)_1^p$ and $U(2)_{-1, -4p-2} \times U(1)_1^{p+2}$, related by particle-hole conjugation. This is the foundation for the Jain-SC transitions in the main text and in SM Sec.~\ref{app:JainSC_transition}. 

\section{A family of superconductors described by \texorpdfstring{$U(2)$}{} gauge theories}\label{app:U(2)SC}

In this section, we briefly review a family of $U(2)$ Chern-Simons theories that can describe superconductors with interesting topological properties. Remarkably, as shown in the main text and in SM Sec.~\ref{app:JainSC_transition}, each member in this family is connected to $\mathrm{Jain}_p$ for some choice of $p$ through a direct quantum phase transition. 

The most general $U(2)$ Chern-Simons theory is labeled as $U(2)_{k_1,k_2}$. Readers who are interested in a detailed pedagogical construction of this theory can refer to Appendix A of Ref.~\cite{Shi2025_dopeMR}. For the present discussion, we will not treat the general case and instead focus on a special class of theories $U(2)_{2k,0}$ with $k \in \mathbb{Z}$. The Lagrangian for this theory takes the form
\begin{equation}
    L_{U(2)_{2k,0}} = - \frac{2k}{4\pi} \Tr \left[\alpha \, d \alpha + \frac{2}{3} \alpha^3\right] + \frac{k}{4\pi} \Tr \alpha \, d \Tr \alpha + \frac{n}{2\pi} \Tr \alpha \, d A \,,
\end{equation}
where $\alpha$ is a $U(2)$ gauge field and $A$ is the external electromagnetic gauge field. Decomposing $\alpha$ into its $SU(2)$ and $U(1)$ components $\alpha = \alpha_0 I + \sum_{a=1}^3 \alpha_a \sigma^a/2$, one can easily verify that the non-Abelian $SU(2)$ Chern-Simons level is $2k$ and the Abelian $U(1)$ Chern-Simons level is $0$~\cite{Witten1989_CSJones}. This level assignment explains the notation $U(2)_{2k,0}$. Since we will be interested in effective field theories that can emerge from microscopic electronic lattice models, we will always choose $A$ to be a $\mathrm{spin}_{\mathbb{C}}$ connection whose charge-1 source is the microscopic electron. As we will see, this condition puts constraints on the allowed values of $k$ and $n$.

To elucidate the structure of this theory, we follow the notation of Ref.~\cite{Seiberg2016_TQFTgappedbdry} and write the $U(2)_{k_1, k_2}$ as
\begin{equation}
    U(2)_{k_1, k_2} = \frac{SU(2)_{k_1} \times U(1)_{k_2}}{\mathbb{Z}_2} \,, 
\end{equation}
where the $\mathbb{Z}_2$ quotient is implemented by gauging the $\mathbb{Z}_2$ 1-form symmetry in the factorized TQFT $SU(2)_{k_1} \times U(1)_{k_2}$ generated by a Wilson line that carries the $j = k_1/2$ irrep of $SU(2)$ and the $n = k_2/2$ irrep of $U(1)$. In the special case $k_1 = 2k, k_2 = 0$, the $\mathbb{Z}_2$ quotient acts only on the $SU(2)$ sector and reduces the topological order $SU(2)_{2k}$ to $SU(2)_{2k}/\mathbb{Z}_2$. The endpoint of the Wilson line with $j = k$ is an anyon in $SU(2)_{2k}$ with topological spin $k(k+1)/(2k+2) = k/2$. When $k$ is odd/even, this anyon has fermionic/bosonic self-statistics and the quotient theory $SU(2)_{2k}/\mathbb{Z}_2$ is a fermionic/bosonic topological order. 

Now let us bring back the $U(1)$ sector with Chern-Simons level $k_2 = 0$. With a vanishing Chern-Simons level, the dynamics of $\Tr \alpha$ is dominated by the conventional Maxwell term. Importantly, since the system has a global $U(1)$ symmetry generated by the background gauge field $A$, monopoles of $\Tr \alpha$ are prohibited in the Lagrangian and the $U(1)$ Maxwell theory does not confine at low energy. The gapless photon of this Maxwell theory is precisely the particle-vortex dual of the Goldstone mode in a superconductor (with $A$ treated as a background non-dynamical field). As shown in Ref.~\cite{Shi2026_4eTQC}, the flux quantization of this superconductor is in units of $h/(2ne)$, where $n$ is the coefficient appearing in the mutual Chern-Simons coupling between $\Tr \alpha$ and $A$. Gluing the $U(1)$ sector back to the $SU(2)$ sector, we therefore conclude that $U(2)_{2k,0}$ realizes an SC* state in which a charge-$2ne$ superconductor coexists with a $SU(2)_{2k}/\mathbb{Z}_2$ topological order. 

Finally, let us return to the relationship between $k$ and $n$. From the earlier discussion, we know that $SU(2)_{2k}/\mathbb{Z}_2$ is a fermionic/bosonic topological order for odd/even $k$. Moreover, the anyon carrying the spin-$k$ irreducible representation of $SU(2)_{2k}$ becomes a local fermionic/bosonic excitation for odd/even $k$ in $SU(2)_{2k}/\mathbb{Z}_2$. In the $U(2)$ gauge theory, the local operator carrying the spin-$k$ representation of $SU(2)$ is precisely the elementary monopole operator $\mathcal{M}_{\Tr \alpha}$ which inserts $2\pi$ flux of $\Tr \alpha$~\cite{Shi2025_dopeMR}. Therefore, the monopole $\mathcal{M}_{\Tr \alpha}$ is identified with a microscopic local fermion for odd $k$ and a local boson for even $k$. On the other hand, the mutual Chern-Simons coupling between $\Tr \alpha$ and $A$ implies that the same monopole $\mathcal{M}_{\Tr \alpha}$ also carries physical electric charge $n$. In order for this charge assignment to be consistent, $n$ and $k$ must have the same parity. One can check that this parity matching condition is indeed satisfied for all the superconductors constructed in the main text and in SM Sec.~\ref{app:JainSC_transition}.

\section{Bandwidth-tuned Jain-SC transition at general filling \texorpdfstring{$\nu = p/(2p+1)$}{}}\label{app:JainSC_transition}

In this section, we generalize the ideas in the main text to construct phase transitions from the Jain FCI at $\nu = p/(2p+1)$ to several distinct superconductors. For generic values of $p$, the superconductors that we find will not be simple charge-2e topological superconductors, but rather higher charge superconductors that generically coexist with a gapped topological order. 

\subsection{Superconductivity from the Abelian representation of \texorpdfstring{$\mathrm{Jain}_p$}{}}

Let us first consider the simplest Abelian representation $L_{U(1) \times U(1), p}$ of $\mathrm{Jain}_p$. This description can be obtained by fractionalizing the electron as $c = f_1 f_2 f_3$ and putting $f_i$ in Chern insulators with $C_1 = C_2 = 1, C_3 = p$. Though the Chern numbers for $f_1, f_2$ are the same, we choose their wavefunctions to be distinct so that the ansatz preserves the gauge group $U(1) \times U(1)$. 

Following the logic in the main text, we focus on phase transitions in which the fermionic partons go through band-inversions. This class of transitions are particularly simple as they are free in the mean-field approximation and do not require additional layers of fractionalization. Due to the fractionalization of translation symmetry, we know that lattice translations act projectively on the partons, with a projective phase $T_x T_y T_x^{-1} T_y^{-1} = e^{2\pi i q/(2p+1)}$, where $q$ is coprime with $2p+1$. As a result, the momentum of each parton is only well-defined in a $|2p+1|$-fold reduced Brillouin zone, and the dispersion of each parton is $|2p+1|$-fold degenerate within the reduced Brillouin zone~\cite{Cheng2016_SET_translation,Shi2024_doping}. This degeneracy implies that across a band-inversion transition, the Chern number of each parton can only change by $\Delta C_i = \pm (2p+1)$. Furthermore, superconductivity requires the resistivity tensor to vanish. By the Ioffe-Larkin rule~\cite{Ioffe1989_rule}, this condition translates to
\begin{equation}
    R = \frac{h}{e^2} \sum_i C_i^{-1} \begin{pmatrix}
        0 & 1 \\ -1 & 0 
    \end{pmatrix} = 0 \quad \rightarrow \quad \sum_i C_i^{-1} = 0 \,. 
\end{equation}

Given these constraints, we can ask: by changing one of the Chern numbers, could we transition from $\mathrm{Jain}_p$ to a superconductor? This question can be answered by a simple casework. 
\begin{enumerate}
    \item If $C_1 \rightarrow C_1 + 2p+1$ and $C_2, C_3$ stay fixed, then 
    \begin{equation}
        \sum_i C_i^{-1} =(2p+2)^{-1} + 1 + p^{-1} = \frac{p + 2(p+1)p+2(p+1)}{2(p+1) p} = \frac{2p^2+5p+2}{2p(p+1)} \,. 
    \end{equation}
    The only solution to $R = 0$ with integer $p$ is therefore $p = -2$. 
    \item If $C_1 \rightarrow C_1 - (2p+1)$ and $C_2, C_3$ stay fixed, then 
    \begin{equation}
        \sum_i C_i^{-1} = (-2p)^{-1} + 1 + p^{-1} = \frac{-1 + 2p + 2}{2p} = \frac{2p+1}{2p} \,. 
    \end{equation}
    The equation $R = 0$ now has no integer solution.
    \item If $C_1, C_2$ are fixed and $C_3 \rightarrow C_3 + 2p+1$, then 
    \begin{equation}
        \sum_i C_i^{-1} = 1 + 1 + (3p+1)^{-1} = \frac{2(3p+1) + 1}{3p+1} = \frac{6p + 3}{3p+1} \,. 
    \end{equation}
    The equation $R = 0$ again has no integer solution.
    \item If $C_1, C_2$ are fixed and $C_3 \rightarrow C_3 - (2p+1)$, then 
    \begin{equation}
        \sum_i C_i^{-1} = 1 + 1 - (p+1)^{-1} = \frac{2p+2 - 1}{p+1} = \frac{2p+1}{p+1} \,. 
    \end{equation}
    The equation $R = 0$ again has no integer solution.
\end{enumerate}
From this simple casework, we see that within the Abelian gauge theory representation, the only possible parton band-inversion transition between $\mathrm{Jain}_p$ and a superconductor is the transition between $\mathrm{Jain}_{-2}$ ($\nu = 2/3$) and the charge-2e SC with chiral central charge $c_- = -2$. For all other values of $p$, we need the more complicated machinery of $U(2)$ gauge theory, to which we now turn. 

\subsection{Superconductivity from non-Abelian representations of \texorpdfstring{$\mathrm{Jain}_p$}{}}

Let us now return to the parton construction of $\mathrm{Jain}_p$ in which $c = f_1 f_2 f_3$ with $C_1 = C_2=1, C_3 = p$ and $f_1, f_2$ having identical mean-field solutions. This ansatz preserves a $U(2)$ gauge structure under which $(f_1, f_2)$ transforms in the fundamental representation. Across a band-inversion transition, we therefore demand that $f_1, f_2$ change their Chern numbers simultaneously by $\Delta C_1 = \Delta C_2 = \pm (2p+1)$. The unique choice that gives superconductivity is $\Delta C_1 = \Delta C_2 = - (2p+1)$, which leads to $C_1 = C_2 = - 2p$. Integrating out the gapped partons gives the TQFT
\begin{equation}
    \begin{aligned}
    L^{(+), \rm SC}_{U(2), p} &= - \frac{2p}{4\pi} \Tr \left[(\alpha + AI) d (\alpha + AI) + \frac{2}{3} \alpha^3\right] - 8 p \mathrm{CS}_g + \frac{p}{4\pi} (\Tr \alpha + A) \, d (\Tr \alpha + A) + 2 p \mathrm{CS}_g \\
    &= - \frac{2p}{4\pi} \Tr \left[\alpha \,d \alpha + \frac{2}{3} \alpha^3\right] + \frac{p}{4\pi} \Tr \alpha \, d \Tr \alpha - \frac{p}{2\pi} \Tr \alpha \, d A - 3 p \mathrm{CS}[A,g] \,.
    \end{aligned}
\end{equation}
The last line describes a $U(2)_{2p, 0} \times U(1)_{-1}^{3p}$ theory. Since the $U(1)$ level of $U(2)_{2p,0}$ vanishes, this phase is indeed a superconductor with a gapless Goldstone mode described by the Maxwell term of $\Tr \alpha$. The mutual Chern-Simons coupling between $\alpha$ and $A$ means that this phase is a charge-$|2p|$ superconductor with flux quantization in units of $h/(2pe)$~\cite{Shi2026_4eTQC}. The chiral central charge of this general superconductor is given by 
\begin{equation}
    c_- = \sgn(p) \frac{3 (2 |p|)}{2|p|+2} - 3p = \begin{cases} \frac{3p - 3p(p+1)}{p+1} = - \frac{3p^2}{p+1} & p > 0 \\ \frac{3 p - 3p (1-p)}{1-p} = \frac{3p^2}{1-p} & p < 0 \,. \end{cases}
\end{equation}
When $p = 1$, the $U(2)_{2,0} \times U(1)_{-1}^3$ theory describes a charge-2 chiral topological superconductor with $c_- = 3/2 - 3 = -3/2$. For more general values of $p$, the $U(2)_{2p,0}$ sector endows the superconductor with a residual topological order $SU(2)_{2p}/\mathbb{Z}_2$, which is generically non-Abelian. 

The critical theory for this transition contains $|2p+1|$ gapless Dirac fermions in the fundamental representation of $U(2)$, coupled to the background Chern-Simons terms in $L^{(+)}_{U(2),p}$. The minimally charged anyon that goes gapless across this transition is the anyon sourced by $f_1$ or $f_2$, which carries a fractional charge $p/(2p+1)$. Therefore, the minimal local excitation that goes gapless is a charge-$p$ composite particle. When $p$ is odd/even, the critical field theory is fermionic/bosonic respectively. 

Following the logic in the main text, we can obtain a different class of Jain-SC transitions by viewing $\mathrm{Jain}_p$ as the particle-hole conjugate of $\mathrm{Jain}_{\tilde p}$ with $\tilde p = - (p+1)$. In other words, we stack an IQH state at $\nu = 1$ with a $\mathrm{Jain}_{\tilde p}$ state of the additional electrons at filling $\nu = \tilde p/(2\tilde p+ 1)$. The corresponding parton construction is $c = f_1 f_2 f_3$ with $C_1 = C_2 = 1$ and $C_3 = \tilde p = - (p+1)$. The only way to transition to a superconductor through parton band-inversion is to change the Chern numbers of $f_1$ and $f_2$ to $C_1 = C_2 = -2 \tilde p = 2(p+1)$. The resulting TQFT is 
\begin{equation}
    \begin{aligned}
        L^{(-, SC)}_{U(2), p} &= \mathrm{CS}[A,g] - L^{(+, SC)}_{U(2), -(p+1)} \\
        &= - \frac{2(p+1)}{4\pi} \Tr \left[\alpha d \alpha + \frac{2}{3} \alpha^3\right] + \frac{p+1}{4\pi} \Tr \alpha\, d \Tr \alpha - \frac{p+1}{2\pi} \Tr \alpha \, d A - (3p+2) \mathrm{CS}[A,g] \,.
    \end{aligned}
\end{equation}
We recognize this theory as $U(2)_{2(p+1), 0} \times U(1)_{-1}^{3p+2}$, which is a charge-$|2(p+1)|$ chiral superconductor with coexisting topological order $SU(2)_{2(p+1)}/\mathbb{Z}_2$. Setting $p = -2$ gives a direct transition from the Jain state at $\nu = 2/3$ to a charge-2 chiral topological superconductor with $c_- = 5/2$, which agrees with our analysis in the main text. 

The critical theory for this transition again contains $|2p+1|$ gapless Dirac fermions in the fundamental representation of $U(2)$, coupled to $L^{(-)}_{U(2),p}$. The minimally charged anyon that goes gapless across this transition is the anyon sourced by $f_1$ or $f_2$, which carries a fractional charge $\tilde p/(2\tilde p+1) = \frac{p+1}{2p+1}$. Therefore, the minimal local excitation that goes gapless is a charge-$|p+1|$ composite particle. When $p$ is odd/even, the critical field theory is bosonic/fermionic respectively. Note that this pattern is opposite from the previous family of phase transitions based on $L^{(+)}_{U(2),p}$. 

\begin{comment} 
\section{Critical Lagrangian and hierarchy of energy gaps}

To elucidate the nature of this family of Jain-SC transitions, let us study the critical Lagrangian and track the evolution of gauge-invariant observables across each phase transition. The basic gauge-invariant observable is the hierarchy of anyon energy gaps. For a general Jain state, let us label the unique anyon with fractional charge $q$ as $a_q$. In the $L_{p,+}$ representation, $f_1, f_2$ source the $a_{p/(2p+1)}$ anyon while $\psi$ sources the $a_{1/(2p+1)}$ anyon. Approaching the Jain-SC transition, the energy gaps of $f_1$ and $f_2$ close continuously while the energy gap of $\psi$ stays finite. Translating to physical anyon gaps, we conclude that the gap to $a_{p/(2p+1)}$ vanishes while the gap to $a_{1/(2p+1)}$ stays open. In fact, assuming the stability of single-anyon excitations away from the critical point, all anyons that can be generated from $a_{p/(2p+1)}$ through fusion close their gaps simultaneously. In the generic case $p \neq 1$, $a_{1/(2p+1)}$ cannot be obtained from $a_{p/(2p+1)}$ through fusion and all single-anyon states in the $a_{1/(2p+1)}$ sector remain gapped across the phase transition. In fact, among all anyons with charge $q < 1$, only $a_{p/(2p+1)}$ and $a_{2p/(2p+1)}$ are gapless. 
\end{comment}

\section{\texorpdfstring{$U(2)$}{}-symmetric transition between a boson IQH state and a boson Mott insulator}\label{app:BIQH_transition}

Let us consider a lattice model of spinful bosons $\Phi_{\sigma}$ with a global $U(2)$ symmetry acting as $\Phi_{\sigma} \rightarrow U_{\sigma \sigma'} \Phi_{\sigma'}$ such that each spin species sees $2\pi$ magnetic flux per unit cell. At this filling, there are two simple invertible candidate ground states: (1) a topologically trivial bosonic Mott insulator; (2) a bosonic integer quantum Hall (bIQH) state. It is interesting to ask: does there exist a direct transition between these two phases? If the global symmetry $U(2)$ is explicitly broken down to $U(1) \times U(1)$, it is known that a direct transition exists only in the presence of additional lattice symmetries (e.g. a $\mathbb{Z}_2$ inversion symmetry)~\cite{Grover2012_bIQHtransition}. In this section, we show that with the full $U(2)$ global symmetry, additional lattice symmetries are no longer necessary to protect the direct transition. We will first give a parton construction of the bIQH state and use it to derive the critical Lagrangian for the phase transition. 

To construct the bIQH state, we use a parton decomposition $\Phi_{\sigma} = f_{\sigma} \psi$. This decomposition introduces a $U(1)$ gauge redundancy which we remove using a dynamical $\mathrm{spin}_{\mathbb{C}}$ connection $a$ that couples with opposite signs to the fermions $f_{\sigma}$ and $\psi$
\begin{equation}
    L[\Phi_{\sigma}, \alpha] = L[f_{\sigma}, \alpha - a] + L[\psi, a] \,.
\end{equation}
Here, $\alpha$ is a background $U(2)$ gauge field and $\Phi_{\sigma}$ transforms in the fundamental representation of $U(2)$. Writing the non-Abelian gauge field as $\alpha = \alpha_0 I + \sum_a \alpha_a \sigma^a$, we know that $\alpha_0$ has $2\pi$ flux per unit cell. Let us now consider a mean-field ansatz in which $a$ has $4\pi$ flux per unit cell. In this flux configuration, $\psi$ is at filling 1 relative to the magnetic field it sees and forms a Chern insulator with $C = 1$. Each spin species $f_{\sigma}$ is at filling $-1$ relative to the magnetic field it sees and forms a Chern insulator with $C = -1$. Integrating out the partons generates an effective topological field theory
\begin{equation}
    L = - \frac{1}{4\pi} \Tr \left[(\alpha - a) d (\alpha - a) + \frac{2}{3} (\alpha - a)^3 \right] + \mathrm{CS}[a,g] = - \frac{1}{4\pi} \Tr \left[\alpha d \alpha + \frac{2}{3} \alpha^3 \right] + \frac{1}{2\pi} \Tr \alpha \, d a - \mathrm{CS}[a,g] \,. 
\end{equation}
Integrating out $a$ gives 
\begin{equation}
    L = - \frac{1}{4\pi} \Tr \left[\alpha d \alpha + \frac{2}{3} \alpha^3 \right] + \frac{1}{4\pi} \Tr \alpha \, d \Tr \alpha \,. 
\end{equation}
To see that this response theory describes the bIQH state, we can explicitly break the $U(2)$ symmetry down to its maximal Abelian subgroup $U(1) \times U(1)$. In other words, we restrict to gauge field configurations parametrized by $\alpha = \mathrm{diag}(\alpha_1, \alpha_2)$. With this parametrization, the Lagrangian simplifies to
\begin{equation}
    L_{U(1) \times U(1)} \rightarrow - \frac{1}{4\pi} (\alpha_1 d \alpha_1 + \alpha_2 d \alpha_2) + \frac{1}{4\pi}(\alpha_1 + \alpha_2) d (\alpha_1 + \alpha_2) = \frac{1}{2\pi} \alpha_1 d \alpha_2 \,,
\end{equation}
which is precisely the response theory of the $U(1) \times U(1)$ symmetric bIQH state~\cite{Levin2013_bIQH}. 

At the same lattice filling, it is also possible for $f_{\sigma}$ to go into a $C = 0$ topologically trivial band insulator while preserving the $U(2)$ symmetry. The resulting response theory is
\begin{equation}
    L[\Phi_{\sigma}, \alpha] = 0 + \frac{1}{4\pi} a da + 2 \mathrm{CS}_g \,. 
\end{equation}
This response theory is completely independent of $\alpha$. Moreover, integrating over the fluctuating $a$ corresponds to gauging a fermionic integer quantum Hall state, which leads to a topologically trivial theory. Therefore, we conclude that this state is a trivial bosonic Mott insulator. 

The phase transition from the trivial bosonic Mott insulator to the bIQH state is therefore driven by the simplest $U(2)$-symmetric band inversion transition for the $f_{\sigma}$ partons. The critical Lagrangian takes the form 
\begin{equation}
    L_{\mathrm{bIQH-triv}} = \frac{1}{4\pi} a da + 2 \mathrm{CS}_g + \bar \chi i\slashed{D}_{\alpha - a} \chi + M \bar \chi \, \chi + \ldots \,,
\end{equation}
where the Dirac fermion $\chi$ is regularized so that $M < 0$ gives $C_{f_{\sigma}} = -1$ and $M > 0$ gives $C_{f_{\sigma}} = 0$.

\end{document}